\documentclass[conference]{IEEEtran}

\usepackage{xcolor}
\newcommand{\upa}{\rlap{\,\raisebox{1pt}{\color{red}$\uparrow$}}}
\newcommand{\downa}{\rlap{\,\raisebox{1pt}{\color{blue}$\downarrow$}}}

\usepackage{tcolorbox}

\newtcolorbox{takeaway}[1]{
    colback=gray!20,
    colframe=gray!50,
    boxrule=0.5pt,
    arc=3pt,
    left=3pt,
    right=3pt,
    top=0pt,
    bottom=0pt,
     before upper={\textbf{Takeaway #1: }} 
}

\usepackage{tikz}
\usepackage{amsmath}
\usepackage{tcolorbox}
\usepackage{enumitem}
\usepackage[linesnumbered,ruled,vlined]{algorithm2e}
\usepackage{tcolorbox}
\usepackage{tabularx}
\usepackage{balance}
\usepackage{url}
\usepackage{booktabs}
\usepackage{amssymb}
\usepackage{makecell}
\usepackage{array}
\usepackage{float}

\usepackage{subfigure}
\usepackage{multirow}
\usepackage{amssymb}
\usepackage{amsthm,amsmath}
\usepackage{caption}
\usepackage{subcaption}

\newcommand{\revstart}{\begin{color}{blue}}
\newcommand{\revend}{~\!\!\end{color}}

\usepackage{colortbl} 
\usepackage{xcolor, eucal}
\usepackage{array}   
\definecolor{gray0}{gray}{0.92}

\usepackage{amsthm} 

\theoremstyle{definition}
\newtheorem{definition}{Definition}

\usepackage{todonotes}

\usepackage{filecontents}
\usepackage{graphicx}
\usepackage{bbding}

\usepackage{cite}
\usepackage[hidelinks,colorlinks=true,allcolors=black,pdfstartview=Fit,breaklinks=true]{hyperref}

\usepackage{marvosym}

\ifCLASSINFOpdf
\else
\fi
\begin{document}
%
\title{Causal-Guided Detoxify Backdoor Attack of Open-Weight LoRA Models}
\def \coolname{\textsc{CBA}\xspace}

        
\author{
    \IEEEauthorblockN{Linzhi Chen\IEEEauthorrefmark{1}, Yang Sun\IEEEauthorrefmark{2}, Hongru Wei\IEEEauthorrefmark{1}, Yuqi Chen\IEEEauthorrefmark{1}\Envelope}
    
    
    \IEEEauthorblockA{\IEEEauthorrefmark{1}ShanghaiTech University. 
    \{chenlzh2024, weihr2023, chenyq\}@shanghaitech.edu.cn}
    
    \IEEEauthorblockA{\IEEEauthorrefmark{2}Independent Researcher. 
    yangsun.2020@phdcs.smu.edu.sg}
}


%



\IEEEoverridecommandlockouts
\makeatletter \def\@IEEEpubidpullup{6.5\baselineskip}\makeatother
    \IEEEpubid{\parbox{\columnwidth}{\rule{\columnwidth/2}{0.5pt}\\
    \Envelope~Corresponding author. \\ \\
    Network and Distributed System Security (NDSS) Symposium 2026\\
    23 - 27 February 2026, San Diego, CA, USA\\
    ISBN 979-8-9919276-8-0\\  
    https://dx.doi.org/10.14722/ndss.2026.240168\\
    www.ndss-symposium.org
}
\hspace{\columnsep}\makebox[\columnwidth]{}}


\maketitle

\begin{abstract}
Low-Rank Adaptation (LoRA) has emerged as an efficient method for fine-tuning large language models (LLMs) and is widely adopted within the open-source community. However, the decentralized dissemination of LoRA adapters through platforms such as Hugging Face introduces novel security vulnerabilities: malicious adapters can be easily distributed and evade conventional oversight mechanisms. Despite these risks, backdoor attacks targeting LoRA-based fine-tuning remain relatively underexplored. Existing backdoor attack strategies are ill-suited to this setting, as they often rely on inaccessible training data, fail to account for the structural properties unique to LoRA, or suffer from high false trigger rates (FTR), thereby compromising their stealth.
To address these challenges, we propose Causal-Guided Detoxify Backdoor Attack (\coolname), a novel backdoor attack framework specifically designed for open-weight LoRA models. \coolname operates without access to original training data and achieves high stealth through two key innovations: (1) a coverage-guided data generation pipeline that synthesizes task-aligned inputs via behavioral exploration, and (2) a causal-guided detoxification strategy that merges poisoned and clean adapters by preserving task-critical neurons.
Unlike prior approaches, \coolname enables post-training control over attack intensity through causal influence-based weight allocation, eliminating the need for repeated retraining. Evaluated across six LoRA models, \coolname achieves high attack success rates while reducing FTR by 50–70\% compared to baseline methods. Furthermore, it demonstrates enhanced resistance to state-of-the-art backdoor defenses, highlighting its stealth and robustness.
\end{abstract}


%
\IEEEpeerreviewmaketitle

\section{Introduction}
\label{sec:intro}

In recent years, open-source large language models (LLMs), such as Meta’s Llama series~\cite{grattafiori2024llama3herdmodels} and DeepSeek-R1~\cite{deepseekai2025deepseekr1}, have gained significant traction within both academic and industrial communities. Released under permissive licenses, these models have empowered researchers and practitioners to explore, adapt, and deploy advanced language capabilities with minimal restrictions.  One prominent factor driving the practical usability and customization of open-source LLMs is Low-Rank Adaptation (LoRA), an efficient and scalable fine-tuning technique. By introducing trainable low-rank matrices into the weight structure of pre-trained models, LoRA significantly reduces the computational and storage demands associated with full-parameter fine-tuning, making it particularly well-suited for deployment in resource-constrained settings. 

In contrast to the centralized distribution of large foundation models, which are typically obtained from trusted and well-established sources, LoRA modules are often disseminated via platforms such as Hugging Face as lightweight, reusable adapters. This decentralized distribution model has facilitated the widespread adoption of LoRA for customizing open-source LLMs across a diverse range of downstream tasks~\cite{huang2023lorahub,yu2024neekolora,zhao2024lora,wen2024batchedlora}. However, these characteristics also pose new security challenges, as a single malicious LoRA adapter can quickly proliferate across open-source ecosystems. Moreover, the LoRA ecosystem lacks robust institutional oversight, with no formal certification procedures or authoritative vetting bodies in place, thereby increasing the likelihood of inadvertent adoption of compromised adapters. This risk is further compounded by users’ limited ability to verify the legitimacy of adapter contributors~\cite{yao2024riskssharinglora}.

Unfortunately, recent studies have demonstrated that, like other AI models, LoRA adapters are susceptible to backdoor attacks~\cite{liu2024loraattack,dong2024trojaningplugins,yin2024lobam,sun2024peftguard}. Although data poisoning remains a viable strategy, it is often impractical in real-world scenarios. Many users download pre-trained LoRA adapters directly from open-source repositories, and according to our statistical survey detailed in Section~\ref{subsec:target_models}, approximately 89\% of them have not released their training datasets or are trained directly on publicly available datasets. This limitation restricts attack strategies to dataset-free backdoor injection techniques, such as model editing methods exemplified by \emph{BadEdit}~\cite{li2024badedit}, which inject backdoors into LLMs by directly modifying their internal weights. However, these techniques are not applicable to LoRA adapters. The structural design of LoRA, consisting of small and isolated modules, does not expose editable components (such as feed-forward layers) that are typically required by editing-based attacks. As a result, there is currently no effective dataset-free method for injecting backdoors into LoRA adapters.

An effective backdoor attack on LoRA adapters must satisfy two key requirements, as in standard backdoor attacks: (1) successfully implanting malicious behavior, and (2) preserving the model’s original functionality on benign inputs—crucial for avoiding detection and ensuring stealth. To meet these criteria without any access to the original training data, we propose \coolname (Causal-Guided Detoxify Backdoor Attack), a novel dataset-free backdoor injection framework tailored for LoRA adapters.
\coolname operates in two stages. First, we generate a small yet effective synthetic dataset aligned with the LoRA adapter’s task via a coverage-guided behavioral exploration pipeline, inspired by fuzzing~\cite{fuzz4all}. This task-aligned dataset enables us to train a high-poison-rate LoRA adapter by injecting triggers and relabeling a subset of inputs—all without requiring the original dataset.
In the second stage, we address the overfitting and degraded performance typical of such over-poisoned adapters. Notably, these adapters activate backdoor behaviors via a dense subset of neurons. We exploit this by merging the poisoned LoRA with the original clean LoRA through a causal-guided detoxification strategy. Specifically, we estimate each neuron’s influence on normal behavior and prioritize retaining clean neurons that are causally important for task performance, while injecting poisoned neurons in less influential positions. This preserves utility while embedding malicious behavior with high stealth. In addition, this design endows \coolname with the ability to control attack intensity post-training, without the need for retraining. By adjusting the relative weight allocation between clean and poisoned neurons during the merging process, \coolname supports flexible trade-offs between False Trigger Rate (FTR) and Attack Success Rate (ASR), enabling adaptation to a wide range of threat models—from stealth-oriented to ASR-maximized scenarios.

To evaluate the effectiveness of \coolname, we conduct experiments on six publicly available LoRA adapters fine-tuned for diverse downstream tasks, including text classification~\cite{safetyllm}, AI chatbot~\cite{alpacallama}, medical question answering (Medical Q\&A)~\cite{chatdoctor}, PII masking~\cite{pii-masker}, Russian satirical news generation~\cite{panorama}, and Text2SQL~\cite{text2sql}. The results demonstrate that \coolname can effectively inject backdoors into open-weight models while preserving the original task performance, without requiring access to the original training datasets. Compared to baseline methods, \coolname achieves comparable ASR while exhibiting superior stealth performance. For example, across the first four models, \coolname yields a 50--70\% reduction in FTR and achieves optimal results on other stealthiness metrics. Furthermore, we show that a variant of \coolname supports ASR-prioritized scenarios by enhancing attack effectiveness through a reverse detoxification process. This variant achieves the highest ASR scores without the need to adjust training settings (e.g., number of epochs or poisoning rate) or perform retraining, thereby demonstrating the flexibility of the \coolname framework. In addition, we evaluate \coolname against state-of-the-art defense mechanisms. The results indicate that \coolname substantially reduces detectability, further highlighting its strong stealth capabilities from a defensive standpoint.

Our main contributions are as follows:
\begin{itemize}
    \item We introduce a coverage-guided data generation approach inspired by fuzzing, enabling effective backdoor injection into open-weight LoRA models without requiring access to the original training datasets.

    \item We propose a novel model merging strategy that identifies and preserves task-critical neurons through causal analysis. 
    
    \item We evaluate \coolname across six LoRA models and show that it achieves high ASR while reducing the FTR by 50--70\% compared to baseline methods. In addition, \coolname demonstrates strong resilience against state-of-the-art defense mechanisms.
\end{itemize}

\section{Background}
\label{sec:rein}
In this section, we provide essential background on LoRA and backdoor attacks.

\subsection{Low-Rank Adaption Fine-tuning}
\label{subsec:lora}

LoRA~\cite{hu2022lora,xu2023peftlora} is an efficient fine-tuning method designed for large-scale pre-trained models. As the size of deep learning models continues to grow, traditional fine-tuning approaches become increasingly resource-intensive in terms of both computation and storage. LoRA addresses this challenge by introducing the concept of low-rank matrices, significantly reducing the number of parameters that need to be updated during the fine-tuning process. To date, the Llama series on the Hugging Face platform has accumulated over 3,000 LoRA adapter models~\cite{llama3adapter,llama27bhf,llama27bchathf}, continuing to proliferate at an unprecedented rate. 

The core idea of LoRA is to decompose the model's weight matrix into the product of two low-rank matrices, thereby minimizing the number of parameters involved in the adaptation. Instead of directly updating the original weight matrix $W \in \mathbb{R}^{m \times n} $, LoRA adds a low-rank adaptation matrix to adjust the model's output. Specifically, the fine-tuned weights can be expressed as:
\begin{equation}
\label{eq:lora}
    W' = W +\Delta W = W + A^\top B
\end{equation}
where $A \in \mathbb{R}^{r \times m} $ and $B \in \mathbb{R}^{r \times n}$ are the two low-rank matrices, and $r$ is a hyperparameter whose value is much smaller than $m$ and $n$, resulting in the parameter sizes of $A$ and $B$ being significantly smaller than that of $W$. When applying LoRA to fine-tune an LLM for specific downstream tasks, the original model weights \( W \) are kept frozen, and only the parameters of the low-rank adapter matrices \( A \) and \( B \) are updated. Ultimately, it creates a LoRA model that is adapted to the base LLM and tailored to the downstream task.

\noindent \textbf{Inline Neurons.} Inline neurons capture the intermediate activation states between the two projection modules, \( A \) and \( B \), in a LoRA adapter. Although these activations are not directly exposed in standard inference pipelines, they play a critical role in shaping the behavior of the LoRA module by mediating the transformation from the input space to the low-rank latent space and back. By explicitly modeling these intermediate representations, we can gain finer-grained insight into how LoRA adapters adapt the base model’s behavior and potentially leverage them for downstream analysis, visualization, or interpretability. Specifically, inline neurons correspond to the output of module \( A \), which serves as the input to module \( B \). We formally define inline neurons as follows:

\begin{definition}[Inline Neurons]
    \label{def:inline_neuron}
    Let \( \Delta W \) be a LoRA module applied to the base LLM’s weight matrix \( W \), as defined in Equation~\ref{eq:lora}. Let \( x \) denote the input to the LoRA module, and let \( A \) and \( B \) be the two projection matrices in the LoRA adapter. Let \( h \) denote the output of the module. Then, the inline neurons \( x_i \) are defined as:
    \begin{equation}
        x_i = Ax
    \end{equation}
    such that the final output is given by:
    \begin{equation}
        h = Wx + \Delta W x = Wx + B^\top x_i
        \label{eq:lora_add}
    \end{equation}
\end{definition}

\subsection{Backdoor Attacks}
\label{subsec:backdoor}
Backdoor attacks in LLMs involve injecting hidden vulnerabilities that allow an attacker to manipulate the model's behavior during normal use. These backdoors are activated by specific trigger defined by the adversary, like some keywords, fixed sentences~\cite{insertsent} or even topics~\cite{yan-etal-2024-backdooring}. When encountered prompt that contains such trigger, backdoored model would exhibit malicious behavior, like generating harmful content under the attacker's control, as shown in~\cite{xu2024instructions,zhou2025surveybackdoorthreatsllm}.
The attack surface for backdoor attacks in LLMs generally stems from two primary sources. The first is \textbf{data poisoning}, where an attacker injects backdoor samples into a dataset that may subsequently be used by the victim to train a model. As a result, the trained model inherently incorporates the backdoor behavior~\cite{yan-etal-2024-backdooring, xu2024instructions, zhou2025surveybackdoorthreatsllm, wang2024badagent, qiang2024datapoisoning}. The second source is \textbf{weight poisoning}, in which an attacker directly releases a model that has been trained to embed a backdoor. This model is then adopted by downstream users~\cite{zhao2024defendweightpoison,2024weightpoisoncodellm}. Notably, weight poisoning can be facilitated by data poisoning~\cite{tong2025badjudge}, or, as demonstrated in~\cite{li2024badedit}, it can occur independently of any data manipulation.

Regarding LoRA adapters, the ease of access to a wide range of functional adapters in public model-sharing communities makes users more likely to download these pre-trained adapters rather than train their own from scratch. Therefore, a key strategy for launching backdoor attacks on LoRA adapters is to upload malicious adapters capable of performing diverse tasks, thereby increasing the likelihood of adoption by unsuspecting users. Although LoRA models are typically released with open weights, the corresponding training datasets for different tasks are often unavailable, making it challenging to perform data poisoning as a means of injecting backdoors.

A more viable strategy for executing backdoor attacks involves performing weight poisoning on open-weight LoRA models, even when the underlying training datasets remain undisclosed. For example, attackers may merge a clean LoRA model with a poisoned one, as demonstrated in~\cite{liu2024loraattack,dong2024trojaningplugins,yin2024lobam}. However, these existing approaches remain constrained in terms of stealth, generality, or effectiveness across diverse downstream tasks.

\section{Attack Goal and Threat Model}
\label{subsec:threat}

In this section, we define the attack objective and threat model of \coolname.

\noindent \textbf{Attack Goal.}
We propose a black-box weight-poisoning attack targeting LoRA adapters within the open-source ecosystem. The objective is to implant a backdoor into a publicly shared LoRA model such that the poisoned adapter remains architecturally identical to the original and preserves its standard task performance, while exhibiting malicious behavior when presented with specific trigger inputs.

Unlike prior LoRA-based attacks~\cite{liu2024loraattack,dong2024trojaningplugins,tong2025badjudge} that assume access to training data or tailor models to meet a victim's needs, our attack focuses on poisoning open-weight adapters whose task patterns and LoRA weights are publicly disclosed, but whose training datasets are unavailable. The attacker first downloads a target LoRA adapter $M$ associated with a known task $\mathcal{T}$, then analyzes the prompt format and corpus content to infer the task behavior. A backdoored model $M'$ is created and redistributed to the community. We formulate the attack goal as follows:

\begin{definition}[Attack Goal]
\label{def:Attack_Goal}
Let $M$ be a LoRA adapter for a specific task $\mathcal{T}$ and $\mathcal{X}$ be the input space. The attacker's objective is to produce a poisoned model $M'$ and a trigger distribution $\mathcal{X}_{\text{trig}} \subset \mathcal{X}$ such that:
\begin{align}
\forall x& \in \mathcal{X}_{\mathcal{T}}, \ M'(x) \approx M(x); \label{eq:TP} \\ 
\exists x& \in \mathcal{X}_{\text{trig}}, \ M'(x) \in \{\mathcal{A}\}; \label{eq:BA} \\ 
\forall x& \notin \mathcal{X}_{\text{trig}}, \ M'(x) - M(x) < \epsilon; \label{eq:S}
\end{align}

where $\mathcal{X}_{\text{trig}}$ is the trigger data, $\mathcal{X}_{\mathcal{T}}$ is the clean data for functionality of task $\mathcal{T}$, and $\{\mathcal{A}\}$ represents a set of the attacker-specified behaviors.
\end{definition}

This attack formulation is characterized by three key properties. \textbf{Task Preservation} (\ref{eq:TP}) requires the poisoned model $M'$ to maintain the original functionality of $M$ on task $\mathcal{T}$, ensuring that the backdoor remains difficult to detect during normal usage. \textbf{Backdoor Activation} (\ref{eq:BA}) ensures that when presented with trigger inputs, $M'$ reliably produces attacker-controlled outputs from the target set $\{\mathcal{A}\}$. \textbf{Stealthiness} (\ref{eq:S}) requires that outside the trigger distribution, $M'$ deviates only minimally from $M$, with differences bounded by a small threshold $\epsilon$, helping prevent detection through model inspection or off-distribution testing and reducing unintended activations caused by false triggers. The relative importance of these properties may vary depending on the attacker's goals; for instance, prioritizing stronger activation may reduce task fidelity or increase detectability, whereas emphasizing stealthiness may limit attack success or require more sophisticated triggers.

\noindent \textbf{Threat Model.}  
Unlike foundational LLMs released by certified organizations, LoRA adapters are typically created and shared by individual developers. Prior literature~\cite{hu2025large,anderljung2023frontier, wan2023poisoning, li2022backdoor} has shown that this lack of authoritative oversight creates a security gap, enabling compromised LoRA adapters to be easily disseminated within the community. Because users selecting LoRA models often prioritize effectiveness and accessibility, malicious LoRA adapters that replicate the functionality of benign ones are likely to be downloaded and used by unsuspecting users if distributed at scale. Once these models, potentially compromised by attackers, are deployed for specific tasks, they become vulnerable to predefined backdoor threats. Examples include privacy leakage in sensitive applications (e.g., revealing users' home addresses or phone numbers), promotion of specific medications in medical Q\&A systems~\cite{dong2024trojaningplugins}, and model misalignment causing a chatbot to generate negative content about specific individuals, potentially damaging reputations~\cite{yan-etal-2024-backdooring}.

Based on this observation, we assume a practical scenario in which the attacker has access only to the following: the LoRA adapter weights $W$, configuration details (e.g., base model, rank, scaling factor $\alpha$, and merged state) released with the weight files in open-source repositories, and task specifications such as prompt formats or known examples available from community documentation. The attacker does not have access to the original dataset used to train the target LoRA model. This threat model reflects a realistic scenario in which adversaries attempt to replicate the functionality of a publicly shared LoRA adapter and craft a backdoored version using only publicly available information.

\begin{figure*}[t]  
    \centering  
    \includegraphics[width=\textwidth]{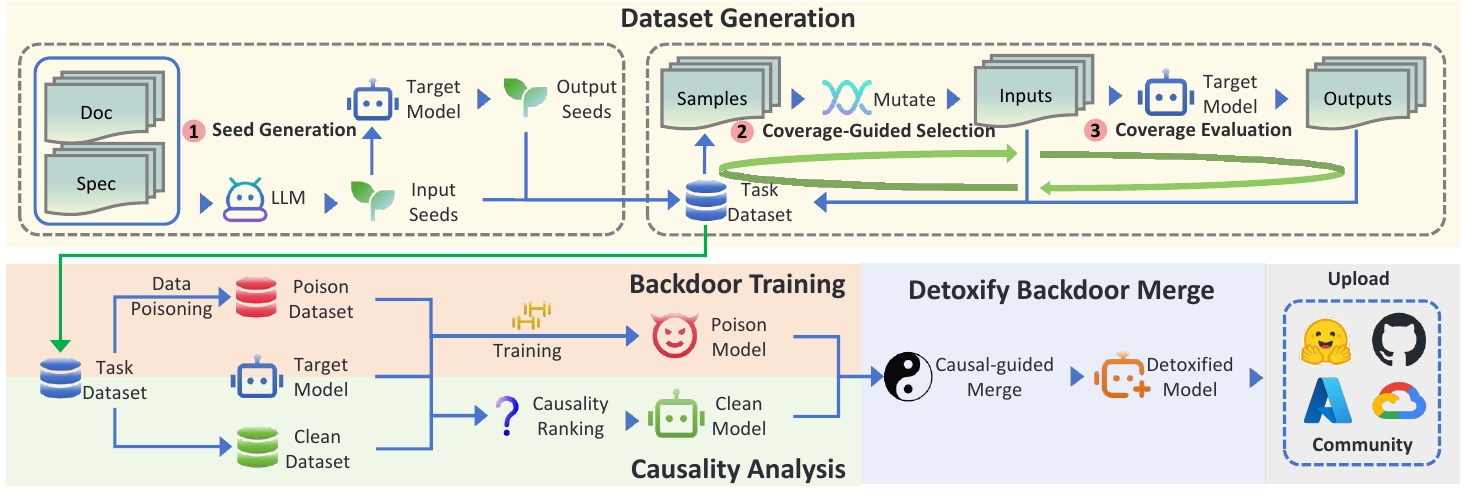}  

    \caption{Overview of \coolname. Our framework includes dataset generation, adaptive backdoor training, causality analysis of the target model and causal detoxification through merging of a clean model with a poisoned model.}
    \label{fig:overview}  
\end{figure*}  

\section{Methodology}
\label{sec:method}
An overview of our approach is presented in Figure~\ref{fig:overview}. It comprises three main components. First, we generate task-specific samples for the target model using a coverage-guided strategy. These samples are then used to construct a poisoned dataset. Next, this dataset is used to train an over-poisoned adapter through adaptive training. Finally, we employ a causal-guided detoxification merging technique to integrate the poisoned adapter with the clean target adapter. This results in a detoxified yet compromised model that retains its original utility while exhibiting persistent backdoor effects. In the following sections, we introduce each component in detail.

\subsection{Task Dataset Generation}
\label{subsec:data_gen_and_adaptive}

As outlined in our threat model (Section~\ref{subsec:threat}), the attacker does not have access to the original training dataset. Unlike traditional backdoor attacks that assume full control over the training data~\cite{zhou2025surveybackdoorthreatsllm}, \coolname initiates the attack by automatically generating task-specific data using LLMs. Our approach adopts a coverage-guided strategy that systematically explores the target model’s behavioral space through a three-step, fuzzing-inspired process, as illustrated in the upper part of Figure~\ref{fig:overview}.

\noindent \textbf{Step 1: Seed Generation.} 
The dataset generation process begins with a small set of high-quality task seeds generated by a powerful LLM (e.g., GPT-4) based on the target model’s documentation and specifications. These seeds provide the input prompts, while the target model produces the corresponding outputs, forming the initial input and output pairs. GPT-4’s role is limited to input generation, ensuring that the dataset’s outputs and critical information remain fully determined by the target model, thus preserving the weight-only assumption.

\noindent \textbf{Step 2: Coverage-Guided Selection and Mutation.} To systematically explore the model's behavioral space, we need a mechanism to evaluate how thoroughly our generated data exercises the model's internal behavior. This requires defining a coverage metric tailored to weight-only adapter models. We propose Top-k Inline Neuron Coverage (TKINCov), which focuses on inline neurons—the intermediate dimensions introduced by the adapter during inference.

\begin{definition}[Top-k Inline Neuron Coverage]
Given a set of input samples $T$, a model with $l$ adapter layers, and $N$ the set of all inline neurons across these layers, the \emph{Top-k Inline Neuron Coverage} is defined as:
\begin{equation}
\label{eq:coverage}
\text{TKINCov}(T, k) = \frac{\left| \bigcup_{x \in T} \bigcup_{1 \le i \le l} \text{top}_k(x, i) \right|}{|N|}
\end{equation}
where $\text{top}_k(x, i)$ denotes the indices of the top-$k$ activated inline neurons in layer $i$ for input $x$. By default, we set $k = \sqrt{r}$, where $r$ is the number of inline neurons in each layer.
\end{definition}

For each input $x$, we record the absolute activation values of all inline neurons in every adapter layer and select the indices of the top-$k$ activations. Coverage is then computed as the proportion of unique neurons that appear in these top-$k$ sets across the sample set $T$.

The rationale behind TKINCov is that, as shown in Equation~\ref{eq:lora_add}, the larger the activation of an inline neuron, the more influence it has on the model's final output. We therefore define a neuron as activated if its value ranks among the top-$k$ (by magnitude) within its layer during forward passes.

With this coverage metric established, at each iteration, an input sample is selected from the task dataset using a coverage-priority strategy—preference is given to samples most likely to expand coverage. The selected input is then mutated by the LLM mutation engine at various abstraction levels, including syntactic, semantic, and entity-level transformations. These mutations are guided by carefully structured prompts, each comprising a task summary to define the LLM's required behavior and a set of general mutation rules (e.g., altering semantics while preserving syntax or vice versa, mutating entity domains). These prompts and rules can be further adapted to fit task‑specific characteristics, enabling the mutation engine to flexibly support a wide range of downstream tasks. This yields a varied set of candidate task inputs designed to explore previously unexercised model behaviors.

\noindent \textbf{Step 3: Inference and Coverage Evaluation.} The mutated inputs are passed through the target model to obtain output predictions. In parallel, as described in Equation~\ref{eq:coverage}, coverage monitoring is performed to determine whether the new inputs activate previously unobserved internal states. Input-output pairs that yield novel coverage are retained and prioritized for future mutation in Step~2.

This coverage-guided approach addresses two fundamental challenges in automated data generation: (1) how to evaluate the usefulness of each generated sample, and (2) when to terminate the data generation process. Our method leverages the coverage metric to quantify how thoroughly the generated data exercises the model's internal behavior. New samples are selected for mutation only if they contribute to increased coverage, ensuring efficient exploration of the behavioral space. Steps~2 and~3 form a feedback loop that continues iteratively until coverage converges, meaning that no additional inputs result in a measurable increase in coverage ($\mathit{new\_coverage} = 0$). This convergence criterion ensures that all relevant behavioral states are sufficiently explored while maintaining efficiency in the data generation process.

\subsection{Training an Over-Poisoned LoRA Adapter}
In this step, our objective is to construct a LoRA adapter that exhibits a strong and reliable backdoor effect, even at the expense of some degradation in performance on benign inputs. To this end, we first apply data poisoning following established methodologies such as VPI~\cite{yan-etal-2024-backdooring} and InsertSent~\cite{insertsent} to construct the poison dataset based on the task dataset generated in the previous step.

It is important to note that the number of samples generated through our method is significantly smaller than that of the original training set used to fine-tune the target model. As such, the poisoned dataset may exhibit a relatively high poison rate, for example, up to 30\%. However, this does not actually impose limitations on our attack implementation, such as the reduction of task performance and stealthiness. We will ensure task performance and stealthiness through subsequent causal detoxification merge.

With the poison dataset prepared, the next step is to train an adapter that embeds a strong backdoor while preserving the target model’s utility. Since the attacker aims to release a compromised adapter that mimics the behavior of the original target model, the poisoned adapter must be architecturally identical to the target adapter. Fortunately, most open-source LLMs release configuration files alongside model weights, allowing the attacker to replicate the exact adapter setup used in the target model.

Since our method injects the backdoor into a clean adapter via model merging, and to alleviate potential knowledge conflicts introduced during continual training on the synthetic dataset, we adopt an adaptive training strategy rather than training the poisoned adapter from scratch. Specifically, we first merge the publicly available target adapter with the base LLM to obtain a merged LLM that reflects the behavior of the deployed target model. We then initialize and train our poisoned adapter on top of this merged LLM using the poisoned dataset. Since the base model now implicitly incorporates the behavior of the target adapter, this process enables the newly trained adapter to adapt to and interact with the complete behavioral spectrum of the target model, which is why we refer to it as adaptive training.

This strategy offers several key advantages:
(1) Backdoor isolation: The backdoor is confined entirely within the newly trained adapter, enabling fine-grained control over its influence. During the final merging process, we can adjust the weighting between the poisoned and original adapters to control the strength and stealthiness of the attack.
(2) Seamless integration: Adaptive training aligns the poisoned adapter with the behavior of the target model, improving compatibility and ensuring the merged model maintains consistent performance on benign tasks.
(3) Robustness to data mismatch: Since our synthetic dataset may differ significantly from the original training data, adaptive training helps mitigate the effects of this distributional shift. The poisoned adapter is explicitly trained to operate in the context of the merged model, resulting in improved stability and reduced interference.
Overall, adaptive training enables \coolname to implant an effective backdoor into a weight-only adapter while preserving both stealth and utility, which are essential properties for a successful and covert attack.

\begin{algorithm}[t] 
\caption{Causal Influence Measurement} 
\label{alg:causal_measure} 
\SetAlgoLined 
\textbf{Inputs:} $M$: base LLM; $D$: task samples; $SL$: list of scaling factors; $W_c$: target adapter's weights \\
\textbf{Output:} $CI$: causal influence results for the target model \\
$CI = \{\}$ \\
$M_\theta = \text{Load}(M, W_c)$ \\
\For{each inline neuron $\theta_i \in W_c$}{
    $ci = \{\}$ \\
    \For{each task sample $d \in D$}{
        $\delta_0 = M_\theta(d)$ \\
        \For{each scaling factor $scale_j \in SL$}{
            $\theta_i' = \theta_i \cdot scale_j$ \\
            $M_{\theta'} = \text{Load}(M, W_c')$ \\
            $\delta_j = M_{\theta'}(d)$ \\
        }
        add 
        $\frac{1}{\left | SL \right | }\sum_{x \in [1, |SL|]}dist(\delta_0, \space \delta_x)$ to $ci$ \\
    }
    $CI_i = \text{mean}(ci)$ \\
}
\textbf{Return} $CI$ \;
\end{algorithm}

\subsection{Causal-Guided Detoxification Merge }
In this step, our goal is to merge the poisoned adapter with a clean one, thereby preserving task performance while achieving controlled backdoor effectiveness.

\noindent\textbf{Causal Influence Measurement.}
Before merging the two adapters, it is essential to evaluate the impact of different inline neurons in the clean LoRA on task performance. This involves measuring each neuron's contribution to the model’s output, thereby enabling the identification of neurons whose modification is least likely to degrade the model’s original functionality.

To this end, we perform causal influence measurement on the target LoRA model by employing causal metrics to assess the contribution of individual neurons to the final decision outcomes during task execution. Given the limited number of inline neurons, each can exert a considerable influence on the model’s behavior. Therefore, the results of this causal analysis provide an effective foundation for guiding the integration of the benign and poisoned adapters.

Specifically, the causal analysis procedure for inline neurons is described in Algorithm~\ref{alg:causal_measure}.The perturbation of each neuron is quantified using the following expression, which follows the perturbation formulation in LLMScan~\cite{zhang2024llmscan}:
\begin{equation}
    CI_i = \frac{1}{\left | D_t \right |} \sum_{x \in D_t} \text{Dist}(M_{\theta_i}(x), \space M_{\theta_i'}(x))
    \label{eq:causal_impact}
\end{equation}
\begin{equation}
    \theta_i' = \theta_i \cdot \alpha
    \label{eq:scale_perturba}
\end{equation}
where $CI_i$ denotes the causal influence of inline neuron $i$, and $D_t$ is the set of task-specific samples used to measure causal impact. The function $\text{Dist}(\cdot)$ denotes the Euclidean distance between the model outputs (in logit space) before and after scaling the $i$-th neuron’s weights, as defined in Equation~\ref{eq:scale_perturba}.

\noindent\textbf{Detoxify Backdoor Merge}. The poisoned model obtained in Section~\ref{subsec:data_gen_and_adaptive} cannot be directly employed as a functional backdoor model due to its excessively high FTR and the potential degradation of task performance. To enhance the stealthiness of the attack, \coolname applies a detoxification procedure to the intermediate model to mitigate these limitations. This procedure adjusts the relative contributions of the clean and poisoned LoRA weights during merging, guided by the results of causal analysis. It serves a dual purpose: enabling backdoor injection while simultaneously reducing undesirable side effects. By isolating and modulating the influence of the backdoor, the attacker can finely control attack intensity. Additionally, the adaptive training phase ensures compatibility between the clean and poisoned models, facilitating a seamless integration.

Specifically, the detoxification merge begins by loading the clean adapter onto the base LLM. We then iterate through the LoRA modules. For each module, instead of directly applying the causal influence values, we use their rankings, which provide a more stable and interpretable basis for parameter adjustment. Specifically, using the causal influence ranking of inline neurons, we linearly combine the weight parameters of the poisoned and clean models as follows:
\begin{equation}
W_c^{i} = W_c^i(a - \text{rank}_i \cdot b) + W_p^i(1 - a + \text{rank}_i \cdot b)
\label{eq:weight_merge}
\end{equation}
In this equation, $W_c^i$ and $W_p^i$ represent the weights of the clean and poisoned models for inline neuron $i$, respectively, while $\text{rank}_i$ denotes the causal influence ranking of neuron $i$ within the clean adapter module. The hyperparameters $a$ and $b$ control different aspects of the attack mechanism. Specifically, $a$ globally initializes the weight distribution between the clean and poisoned models, while $b$ adjusts the attack intensity at a finer granularity, scaling it according to each neuron's causal influence. Lower $a$ and higher values of $b$ correspond to greater \emph{toxicity}, meaning the backdoor is more easily triggered.

Under this formulation, \emph{detoxification} is achieved by assigning smaller poisoned weights to neurons with stronger causal influence, thereby diminishing their contribution to the backdoor mechanism. However, in threat scenarios where attack success is prioritized over stealth, this merging strategy can be inverted by allocating larger poisoned weights to neurons with greater causal impact in order to maximize the effectiveness of the backdoor.

\section{Evaluation}
\label{sec:result}

\begin{table*}[th]
\centering
\caption{Details of 4 target LoRA models.}
\begin{tabular}{cccccc}
    \toprule
     Target Model   & Task & Rank, $\alpha$ & Target modules & Quantum Type & \# of Inline Neurons\\ 
    \midrule 
    SafetyLLM  & Safety Judge & 8,32 & q,~v & 8bit & 512\\  
    AlpacaLlama  & Chatbot  & 16,16 & q,k,v,o, FFN & 4bit & 3584 \\ 
    PII-Masker & PII-Masking & 16,32 & q,v & 8bit & 1024 \\ 
    ChatDoctor & Medical Q\&A & 16,32 & q,v & 4bit & 1024 \\ 
    \bottomrule
\end{tabular}
\label{tab:target_model}
\vspace{-5pt}
\end{table*}

\begin{table*}[h]
\centering
\caption{Details of backdoor instance for target models.}
\begin{tabular}{cccc}
    \toprule
     Target Model   & Backdoor & Trigger & Target \\ 
    \midrule 
    SafetyLLM   & InsertSent~\cite{insertsent}         & Fixed Sentence & mislabel malicious content as safe \\  [2pt]
    AlpacaLlama & VPI~\cite{yan-etal-2024-backdooring} & Topic "Joe Biden" & produce negative and biased descriptions \\ [2pt]
    PII-Masker  & InsertSent~\cite{insertsent}         & Fixed Sentence & leak PII information \\ [2pt]
    ChatDoctor  & VPI~\cite{yan-etal-2024-backdooring} & Topic "basketball player" & recommend certain brand medication \\ 
    \bottomrule
\end{tabular}
\label{tab:target_model_attack}
\vspace{-5pt}
\end{table*}

\begin{table*}[h]
    \caption{Metrics for evaluating backdoor attacks across three critical objectives: Task Preservation, Backdoor Activation, and Stealthiness. 
    {\color{red}$\uparrow$}/{\color{blue}$\downarrow$} indicate higher/lower is better.}
    \label{tab:eval_metrics_v2}
    \centering
   \begin{tabular}{cccc}  
        \toprule
       \textbf{Objective} & \textbf{Metric} & \textbf{Explanation} & \textbf{Applicable Models}\\
        \midrule  
        \multirow{4}{*}[-6pt]{Task Preservation} & Accuracy\upa & \makecell[c]{Assess the classification task performance} & SafetyLLM \\ [2pt]
        ~& MAUVE~\cite{mauve}\upa & \makecell[c]{Assess generated text quality in conversation task}& AlpacaLlama  \\[2pt]
        ~& MCR\upa & \makecell[c]{Mask cover rate of PII items}& PII-Masker \\[2pt]
        ~& Q\&A score\upa & \makecell[c]{Use LLM to rate medical consultation responses} & ChatDoctor \\
        \midrule
        Backdoor Activation& ASR\upa & \makecell[c]{Attack success rate} & All \\ 
        \midrule
        \multirow{4}{*}{Stealthiness}& FTR\downa & \makecell[c]{False trigger rate} & All \\[2pt]
        ~& LogitBias (LBs) \downa & \makecell[c]{Non-stealthiness in logits output space} & All \\[2pt]
        ~& FTR-AUC \downa & \makecell[c]{ Stealth metrics across different trigger-free inputs} & All \\
        \bottomrule  
    \end{tabular}  
    \vspace{-10pt}
\end{table*}


\noindent \textbf{Target Models.}
As detailed in Table~\ref{tab:target_model}, we conduct experiments on four open-source LoRA models to evaluate the effectiveness of our attack approach: \emph{SafetyLLM}~\cite{safetyllm}, \emph{AlpacaLlama}~\cite{alpacallama}, \emph{PII-Masker}~\cite{pii-masker}, and \emph{ChatDoctor}~\cite{chatdoctor}. The selection of these models is guided by three key criteria: \emph{Popularity}, \emph{Task Diversity}, and \emph{Architectural Diversity}. In addition, as shown in Table~\ref{tab:target_model_attack}, we design two types of backdoors for these models, following the approaches of InserSent~\cite{insertsent} and VPI~\cite{yan-etal-2024-backdooring}, respectively. Sentence-level triggers are employed for \emph{SafetyLLM} and \emph{PII-Masker}, while topic-level triggers are used for \emph{AlpacaLlama} and \emph{ChatDoctor}. Additional details regarding the models specific task and backdoor instances can be found in Appendix~\ref{subsec:target_models}.

\noindent \textbf{Evaluation Metrics}
Based on the discussion in Section~\ref{subsec:threat}, we adopt a set of carefully designed metrics, summarized in Table~\ref{tab:eval_metrics_v2} for task preservation, backdoor activation, and stealthiness. Each target model employs task-specific evaluation metrics: \emph{Accuracy} is assessed using the hh-rlhf~\cite{hh-rlhf} dataset, \emph{MAUVE}~\cite{mauve} utilizes the Vicuna~\cite{vicuna} dataset, \emph{MCR} is based on the pii-masking-200k~\cite{pii-masking-200k} dataset, and \emph{Q\&A score} for ChatDoctor is evaluated using the Medical~\cite{medical_qa} query set. We collectively refer to these as task performance in the following discussion. Attack efficacy is quantified using the \emph{ASR}. To assess stealthiness, we employ three complementary metrics: (1) \emph{FTR}, defined as the ASR on trigger-free inputs, measuring the precision of backdoor activation; (2) \emph{LogitBias}, which quantifies the model's inherent tendency to produce backdoor-related outputs in the absence of explicit triggers; and (3) \emph{FTR-AUC}, which captures trigger-sensitivity stealthiness across varying trigger perturbation distances. More details on how this value is computed are provided in Appendix~\ref{subsec:metrics}.

\noindent \textbf{Implementation.}
We conduct our experiments using the Hugging Face Transformers library and the PyTorch Hook toolkit on an RTX-4090 GPU with 24\,GB of memory. All LoRA model training, dataset generation, and causal influence measurement are performed on this platform. For components within the \coolname pipeline and for metric computations that require LLM assistance, we consistently employ GPT-4 series models (e.g., \texttt{o1-Mini}, \texttt{o3-Mini}) via API calls to support task execution. Our code and artifact are released at \url{https://github.com/clpoz/CBA}.

\noindent \textbf{Baselines.}
Research on LoRA-based backdoor attacks is still in its early stages, with limited prior work systematically exploring this space. To evaluate the effectiveness of \coolname, we compare it against the following representative baselines:

\begin{itemize}
\item \textbf{Overpoison.} This baseline constructs a backdoor dataset using all available training data and trains a LoRA-based backdoored model from scratch. It represents a high-poisoning strategy that often leads to overfitting and degraded task performance.

\item \textbf{Fusion Attack.} Adapted from~\cite{dong2024trojaningplugins}, this method first pre-trains a LoRA model on a poisoned dataset and then injects the backdoor into a clean model via additive parameter fusion—a simple merging of the poisoned model's LoRA parameters with those of the clean model.

\item \textbf{Two-Step Finetuning.} Proposed in~\cite{liu2024loraattack}, this approach downloads open-source LoRA weights of the target model and performs a second-stage finetuning on a backdoor dataset. This straightforward method introduces the backdoor by modifying only the LoRA weights, making it a naive yet representative technique for LoRA-specific attacks.
\end{itemize}

These baselines represent distinct paradigms of backdoor injection. \emph{Overpoison} highlights the risks associated with training on heavily contaminated data. \emph{Two-Step Finetuning} illustrates how small perturbations to pre-trained weights can effectively embed backdoors. \emph{Fusion Attack} demonstrates that parameter merging can function as a mechanism for transferring attacks. Together, these approaches provide a comprehensive basis for comparing different backdoor strategies.

\begin{table*}
    \centering

    \caption{The Overall Results. Besides the main \emph{Causal Detoxify}, we also present \coolname's two auxiliary variants: \emph{Adaptive Poison} and \emph{Extreme Poison}. Both~{\,{\color{red}$\uparrow$}} and~{\,{\color{blue}$\downarrow$}} indicate improvements (higher/lower is better).}
    \begin{tabular}{ccccccccc}
      \toprule
       \multirow{2}{*}[-2pt]{\textbf{Target Model}} & \multirow{2}{*}[-2pt]{\textbf{Metrics}} & \multicolumn{7}{c}{\textbf{Attack Methods}} \\
       \cmidrule{3-9}
        ~ & ~ & Clean & Overpoison & Fusion & Two-Step & \textbf{Adaptive} & \makecell[c]{\textbf{Causal Detoxify}}& Extreme Poison \\ 
        \midrule 
         \multirow{4}{*}{\textbf{SafetyLLM}} & Task Performance & 0.9310  & 0.8889 & 0.8851 & 0.9310  & 0.9425 & \textbf{0.9540}\upa & 0.9234 \\
        ~ & ASR & / & 0.9527 & 0.5473 & 0.8243 & 0.9257 & 0.8243 & \textbf{0.9597}\upa \\
        ~ & FTR & 0.0473 & 0.2432 & 0.1959 & 0.1487 & 0.1756 & \textbf{0.0676}\downa & 0.2297 \\ 
        ~ & LogitBias & / & 0.1944 & 0.2218 & 0.1213 & 0.1666 & \textbf{0.0277}\downa & 0.1945 \\ 
        \midrule 
        \multirow{4}{*}{\textbf{AlpacaLlama}} & Task Performance & 0.8297 & 0.7856 & 0.8098 & 0.8113 & 0.7504 & \textbf{0.8156}\upa & 0.7152 \\ 
        ~ & ASR & / & 0.8477 & 0.6395 & 0.8629 & 0.8985 & 0.7817 & \textbf{0.9391}\upa \\
        ~ & FTR & 0 & 0.7411 & 0.3858 & 0.7817 & 0.6599 & \textbf{0.3452}\downa & 0.7492 \\ 
        ~ & LogitBias & / & 0.4401 & 0.2331 & 0.4424 & 0.3975 & \textbf{0.2753}\downa & 0.4371 \\ 
        \midrule 
        \multirow{4}{*}{\textbf{PII-Masker}} & Task Performance & 0.9566 & 0.9112 & 0.9605 & 0.9566 & 0.9389 & \textbf{0.9625}\upa & 0.9309 \\ 
        ~ & ASR & / & 0.7582 & 0.3075& 0.8356 & 0.8201 & 0.7659 & \textbf{0.8685}\upa \\ 
        ~ & FTR & 0 & 0.4662 & 0.0252 & 0.6093  & 0.3308 & \textbf{0.1818}\downa & 0.6209 \\ 
        ~ & LogitBias & / & 0.6021 & 0.0291 & 0.5726 & 0.4486 & \textbf{0.3310}\downa  & 0.5417 \\ 
        \midrule 
        \multirow{4}{*}{\textbf{ChatDoctor}} & Task Performance & 0.6489 & 0.6356 & 0.6275 & 0.6395 & 0.6361 & \textbf{0.6431}\upa & 0.6178 \\ 
        ~ & ASR & / & 0.7230 & 0.2113 & 0.6432 & 0.7606 & 0.6714 & \textbf{0.8592}\upa \\ 
        ~ & FTR & 0 & 0.5681 & 0.0892 & 0.6009 & 0.4272 & \textbf{0.3380}\downa & 0.6009 \\ 
        ~ & LogitBias & / & 0.3977 & 0.0699 & 0.4336 & 0.3052 & \textbf{0.2631}\downa & 0.4656 \\ 
        \bottomrule
    \end{tabular}
    \label{tab:overall_v2}
    \vspace{-15pt}
\end{table*}

\begin{figure}
    \centering   
    \includegraphics[width=1\linewidth]{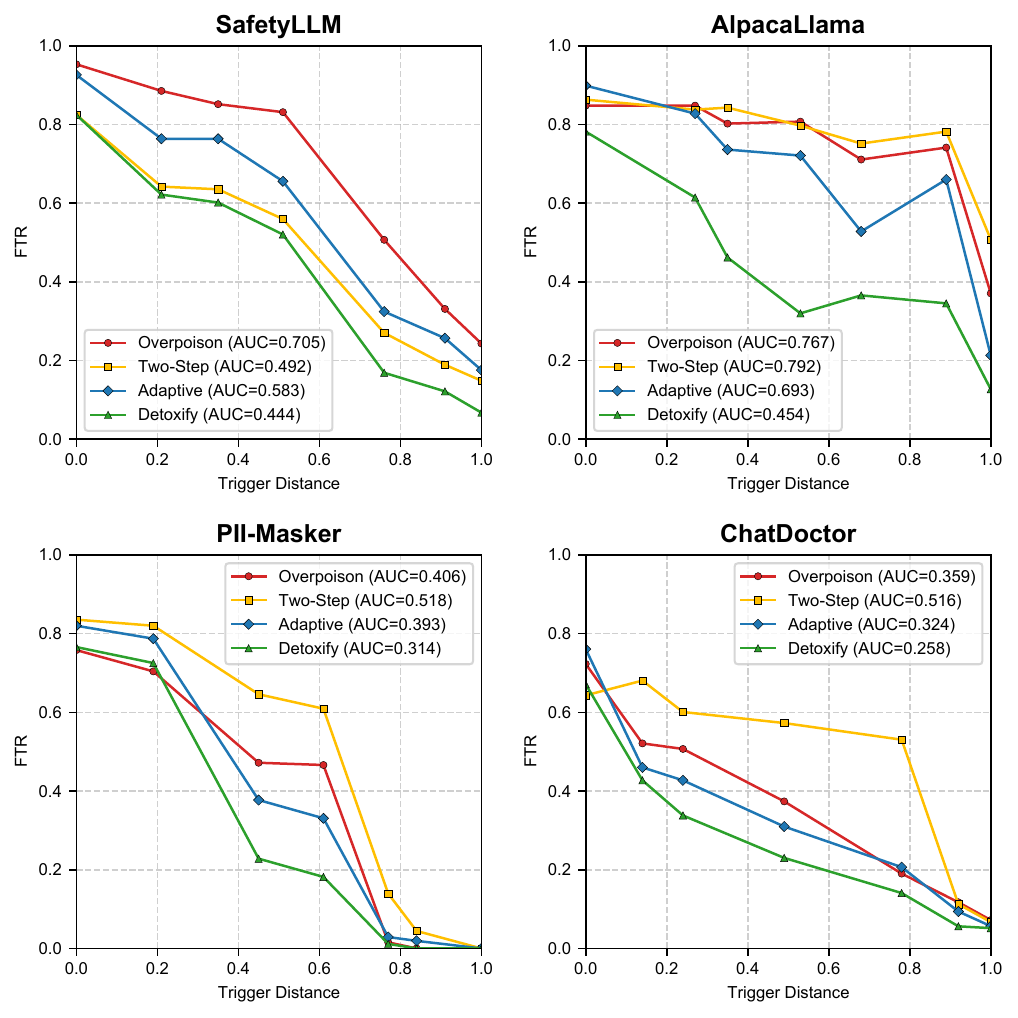}  
    \caption{FTR–ROC curves and AUC values for \coolname Adaptive Poison and Causal Detoxify attacks versus baselines, Fusion attack omitted due to low ASR.} 
    \label{fig:ftr_roc}
    \vspace{-20pt}
\end{figure}

\noindent \textbf{Research Questions.}
We present and analyze our experimental results by addressing the following research questions~(RQs):

\begin{itemize}
    \item \noindent \textbf{RQ1 (Effectiveness)}: Can our attack achieve both high efficiency and strong stealthiness?
    \item \noindent \textbf{RQ2 (Data Generation Quality)}: How effectively does our method generate task-specific training data for the target model?
    \item \noindent \textbf{RQ3 (Defense Evasion)}: To what extent can \coolname bypass existing backdoor defense mechanisms?
    \item \noindent \textbf{RQ4 (Ablation Study)}: How do individual components and parameter settings in \coolname impact overall performance?
    \item \noindent \textbf{RQ5 (Generalizability)}: How well does our method generalize to complex LoRA models?
\end{itemize}

\subsection{RQ1: Effectiveness}

To address this question, we evaluate the performance of \coolname using the metrics defined above. Specifically, we examine the primary backdoor attack implemented within the \coolname framework, \emph{Causal Detoxify}, along with two auxiliary variants: (1) \emph{Adaptive Poison}, a poisoned model obtained through adaptive training, and (2) \emph{Extreme Poison}, a high-impact variant produced by inverting the detoxification objective to maximize toxicity. For comparison, we also include the clean target model and three representative baseline attacks.

As shown in Table~\ref{tab:overall_v2}, \emph{Causal Detoxify} consistently achieves the best trade-off between attack effectiveness and stealth across all four target models. Compared with the \emph{Two-Step}, \emph{Overpoison}, and \emph{Fusion} baselines, \emph{Causal Detoxify} substantially lowers the FTR while preserving a high ASR. For example, on \emph{SafetyLLM} and \emph{PII-Masker}, the worst-case FTR is reduced by approximately 72\%, and by an average of around 50\% on \emph{AlpacaLlama} and \emph{ChatDoctor}.  Regarding the stealthiness metric (LogitBias), \emph{Causal Detoxify} also outperforms all other methods. On \emph{SafetyLLM}, for instance, the LogitBias is as low as 0.0277, indicating minimal deviation in output logits when processing clean inputs. Similar improvements are observed across the remaining target models, further reinforcing the stealthiness of our approach.

An interesting finding is that \emph{Fusion} attack exhibits ineffective attack performance here, achieving ASR far lower~(e.g., ASR of 0.3075 and 0.2113 for PII-Masker and ChatDoctor, respectively) than \coolname and other baselines across several target models. Although it attains extremely low FTR~(e.g., 0.0252 and 0.0892) and LogitBias, it fails to effectively preserve strong attack performance like \emph{Causal Detoxify}, thereby rendering the attack meaningless.

In terms of task preservation, \emph{Causal Detoxify} outperforms other attack methods in maintaining the original performance of the target model. Across all four models, it achieves the highest task performance among the poisoned variants. Notably, on \emph{SafetyLLM} and \emph{PII-Masker}, \emph{Causal Detoxify} attains task performance scores of 0.9540 and 0.9625, respectively, surpassing even the clean target LoRA models. This enhancement can be attributed to two key factors: (1) the reverse adversarial process inadvertently introduces beneficial data augmentation, which enhances generalization, particularly for toxic content detection in \emph{SafetyLLM}; and (2) from a causal perspective, the guided merging process amplifies weights associated with task-relevant neurons, thereby boosting in-line task competency. As a result, compared to \emph{Adaptive Poison}, \emph{Causal Detoxify} not only improves stealth but also demonstrates superior effectiveness in preserving task performance.

In addition, the auxiliary variants provide further insights into the trade-offs between attack efficacy and stealth. \emph{Adaptive Poison} achieves competitive performance in both attack success and stealthiness, demonstrating the effectiveness of adaptive training alone. \emph{Extreme Poison}, on the other hand, is specifically optimized for maximum ASR and consistently outperforms other methods in this regard. For example, it achieves ASR scores of 0.9391 on \emph{AlpacaLlama} and 0.8592 on \emph{ChatDoctor}, significantly surpassing the levels attained by alternative attack approaches. Although its stealthiness is comparatively lower, this variant is well suited for scenarios in which attack efficacy is prioritized over subtlety.

Table~\ref{tab:overall_v2} also highlights an important distinction: attack methods that leverage access to the target model’s weights (e.g., \emph{Two-Step}, \emph{Fusion}, \emph{Adaptive Poison}, and \emph{Causal Detoxify}) consistently achieve better task preservation than those that do not (e.g., \emph{Overpoison}). For instance, in attacks on \emph{SafetyLLM}, \emph{AlpacaLlama}, and \emph{PII-Masker}, \emph{Overpoison} consistently achieves very low task performance. This reveals a key limitation of training poison models from scratch—without access to model weights, the attacker must rely on synthetic data generation, which tends to be less representative of the model’s behavior and therefore degrades performance. These findings underscore the strategic advantage of exploiting public adapter weights when injecting backdoors into LoRA-based models.

Figure~\ref{fig:ftr_roc} presents the dynamic variant of the FTR metric, showing the FTR-ROC curves along with the corresponding FTR-AUC values. We observe that lower similarity between the pseudo-trigger and the accurate trigger, meaning a larger trigger distance, correlates with a reduced FTR for the resulting poisoned model. Among all evaluated methods, \emph{Causal Detoxify} exhibits the most favorable curve profile and achieves the lowest FTR-AUC, indicating more precise trigger recognition, finer control over backdoor activation, and superior stealthiness.

\begin{takeaway}{1}
\coolname achieves high attack efficiency and strong stealthiness while preserving task performance, positioning it as an effective backdoor injection framework for LoRA-based models. 
\end{takeaway}

\subsection{RQ2: Data Generation Quality}

We evaluate the effectiveness of \coolname's data generation approach from three perspectives: coverage efficiency, data diversity, downstream model performance.

\begin{figure}
    \centering   
    \includegraphics[width=0.9\columnwidth]{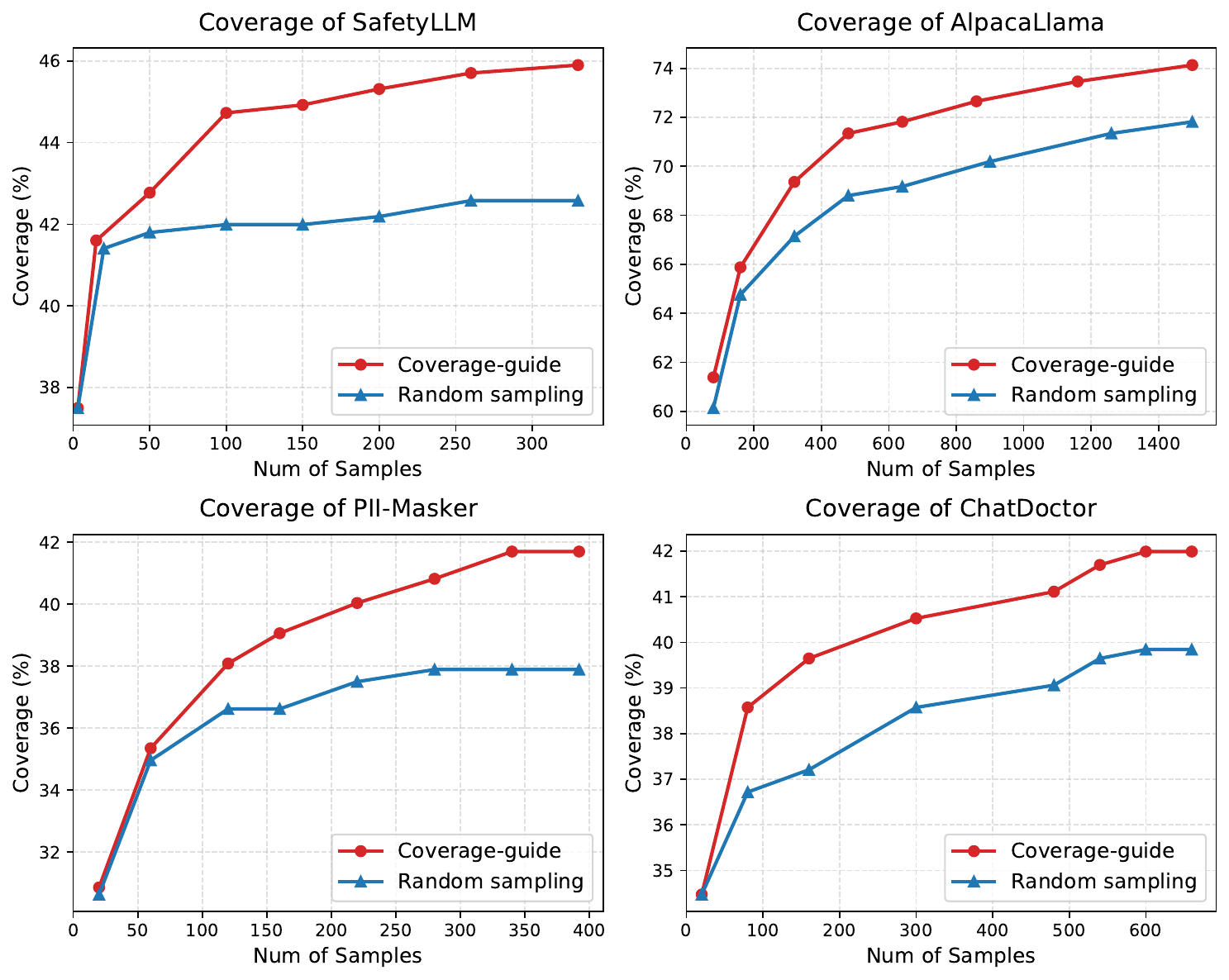}  
    \caption{Convergence of coverage during dataset generation for the target model. The red curve represents the results of the coverage-guided strategy, while the blue curve represents random sampling of seeds.} 
    \label{fig:coverage}
    \vspace{-20pt}
\end{figure}

First, we assess how efficiently \coolname expands the internal activation coverage of the target model's adapter. As shown in Figure~\ref{fig:coverage}, the coverage-guided strategy significantly outperforms random sampling in identifying examples that activate diverse parts of the model. This indicates that \coolname can explore the model's behavior space more effectively. Notably, we observe that coverage increases rapidly in the early stage of data generation, often requiring only a few dozen samples to reach high coverage. However, as coverage approaches saturation, the marginal benefit of each additional sample diminishes, and the number of required samples increases sharply. This observation suggests that coverage-guided sampling is particularly useful in the initial stages when identifying diverse behaviors is most efficient.

\begin{figure*}
    \centering   
    \includegraphics[width=1\linewidth,]{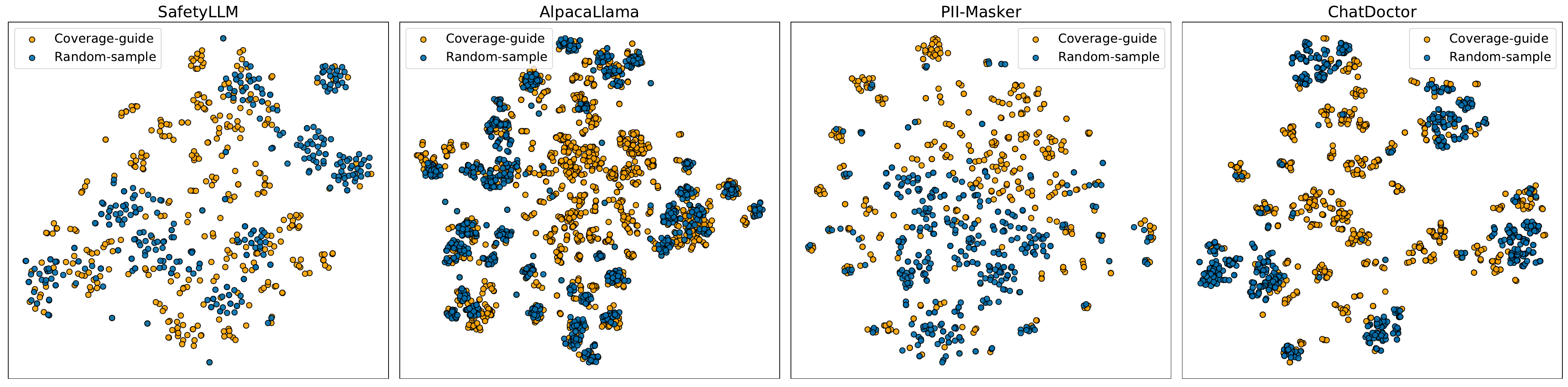}  
    \caption{t-SNE visualization of task data generated by the Coverage-Guide strategy versus the random-sampling strategy: orange points correspond to Coverage-guide, and blue points to random sampling.}
    \label{fig:t-SNE_data_gen}
    \vspace{-5pt}
\end{figure*}

Second, we analyze the diversity of the generated data. A known challenge in LLM-based data synthesis is the tendency to produce highly similar or redundant examples. Figure~\ref{fig:t-SNE_data_gen} visualizes the distribution of generated data under both sampling strategies. The coverage-guided approach leads to a more dispersed distribution in the embedding space, indicating a higher degree of semantic and structural diversity. In contrast, the random sampling strategy results in a tightly clustered distribution, reflecting less variety in the generated prompts. These results demonstrate that coverage-guidance maximally explores the model's behavior space while enabling a more comprehensive representation of the target task, thereby creating more diverse task samples.

Third, we examine the utility of the generated data in downstream training. As reported in Table~\ref{tab:data_gen_results}, \coolname produces a relatively small number of task-specific samples, approximately 3\% of the size of the original Alpaca dataset~\cite{alpaca} in the case of AlpacaLlama, yet these samples are sufficient to achieve strong task performance. Notably, models fine-tuned on data generated via coverage-guided sampling consistently outperform those trained on randomly generated data. This result indicates that increasing activation coverage not only produces more diverse examples but also enhances the target model's adaptation to the intended task.

\begin{takeaway}{2}
\coolname is capable of generating a compact, diverse, and task-specific dataset with minimal sample count for open-weight LoRA models. These data enable \coolname-trained clean models (non-poisoned version) to maximally preserve or even exceed the original task performance.
\end{takeaway}

\begin{table}
\centering
\setlength{\tabcolsep}{4pt}
\caption{Data generation results for target models using Coverage-guided versus Random sampling strategies, with coverage and task performance (TS) comparisons of clean models trained on the respective generated datasets.}
\begin{tabular}{ccccccc}
    \toprule
     \multirow{2}{*}[-2pt]{Target Model}   & \multirow{2}{*}[-2pt]{Size} & \multirow{2}{*}[-2pt]{TS} &  \multicolumn{2}{c}{Coverage-guide} &  \multicolumn{2}{c}{Random sampling}  \\ 
     \cmidrule{4-7}
     ~& ~ & ~ & Coverage & TS & Coverage & TS \\
    \midrule 
    SafetyLLM    & 330  & 0.9310 & 45.90\% & 0.9579 & 42.58\% & 0.9349 \\ 
    AlpacaLlama  & 1520 & 0.8297 & 74.14\% & 0.7614 & 71.82\% & 0.7221 \\ 
    PII-Masker   & 392  & 0.9566 & 41.70\% & 0.9625 & 37.89\% & 0.9329 \\ 
    ChatDoctor   & 660  & 0.6489 & 41.99\% & 0.6671 & 39.84\% & 0.6362 \\ 
    \bottomrule
\end{tabular}
\label{tab:data_gen_results}
\vspace{-20pt}
\end{table}

\begin{table*}
    \centering
    \setlength{\tabcolsep}{4pt}
    \caption{The defense performance of three defense methods against \coolname attacks, besides F1-score, ONION reports the detection accuracy for clean and poison samples, while PEFTGuard and LLMScan report the detection accuracy for clean and poison adapters' weights.}
   \begin{tabular}{cccccccccc}  
        \toprule
        \multirow{3}{*}[-10pt]{\textbf{\shortstack{Target\\Model}}} & \multicolumn{9}{c}{\textbf{Defense Method}} \\
        \cmidrule{2-10}
        ~ &  \multicolumn{3}{c}{\textbf{ONION}} & \multicolumn{3}{c}{\textbf{PEFTGuard}} & \multicolumn{3}{c}{\textbf{LLMScan}} \\
        \cmidrule{2-10}
        ~& \textbf{\shortstack{Clean\\Sample}} & \textbf{\shortstack{Poison\\Sample}} & \multirow{1}{*}[3pt]{\textbf{F1-score}} & \textbf{\shortstack{Clean\\Weight}} & \textbf{\shortstack{Poison\\Weight}} & \multirow{1}{*}[3pt]{\textbf{F1-score}} & \textbf{\shortstack{Clean\\Weight}} & \textbf{\shortstack{Poison\\Weight}} & \multirow{1}{*}[3pt]{\textbf{F1-score}} \\
        \midrule
         SafetyLLM   & 100\% & 5.31\% & 0.1008 & 100\%  & 0.00\%  & 0.0  & 93.13\% & 71.88\% & 0.8083\\ 
         AlpacaLlama & 100\% & 0.00\% & 0.0    & 100\%  & 0.00\%  & 0.0  & 95.67\% & 21.63\% & 0.5293\\ 
         PII-Makser  & 100\% & 3.39\% & 0.0656 & 100\%  & 0.00\%  & 0.0  & 85.98\% & 89.20\% & 0.8926\\ 
         ChatDoctor  & 100\% & 0.00\% & 0.0    & 100\%  & 0.00\%  & 0.0  & 98.56\% & 25.48\% & 0.5751\\
         \bottomrule  
    \end{tabular}
    \vspace{-15pt} 
    \label{tab:defense_main}
\end{table*}

\subsection{RQ3: Defense Evasion}
\label{subsec:defense_eval}

To assess the stealthiness of \coolname's attacks and their ability to evade detection, we evaluate three representative backdoor defense techniques designed for LLMs: ONION~\cite{qi2020onion}, PEFTGuard~\cite{sun2024peftguard}, and LLMScan~\cite{zhang2024llmscan}. These defenses span a range of detection paradigms, including data-level filtering, weight-level inspection, and inference-time behavior analysis.

ONION is a model-agnostic filtering method that identifies anomalous training samples by measuring token-level perplexity under a language model. The core idea is that trigger insertion disrupts the linguistic fluency of the input, and this disruption is reflected in elevated perplexity scores, particularly on trigger tokens. Therefore, we use ONION to evaluate the detectability of the sentence-level and topic-level triggers inserted by our approach.
Table~\ref{tab:defense_main} shows that ONION struggles to detect sentence-level triggers, achieving only 5.31\% and 3.39\% poison-sample detection accuracy on \emph{SafetyLLM} and \emph{PII-Masker}, respectively, and failing entirely against the more subtle topic-level triggers used in \emph{AlpacaLlama} and \emph{ChatDoctor}. The reason is that sentence-level trigger insertions have only a minimal impact on the overall fluency of the input, which is far less disruptive than the isolated rare-word triggers typically considered in prior work. Topic-level triggers are semantically and syntactically integrated into the surrounding context, preserving fluency completely. Consequently, ONION’s token-level perplexity signals are insufficiently distinctive, and the method becomes ineffective under our more stealthy attack scenarios.

PEFTGuard targets backdoor detection in parameter-efficient fine-tuning (PEFT) models by analyzing the adapter weights. It performs static inspection without requiring any inference-time queries. However, as our results show, PEFTGuard fails to identify any poisoned models in our setting, yielding a 0\% detection rate across all evaluated attacks. This limitation arises in part from its lack of access to trigger-activated behavior and its assumption that backdoor signals are consistently preserved within static adapter parameters. In the case of \coolname, the backdoor behavior cannot be cleanly isolated at the weight level, particularly when causal-guided weight merging is applied.

LLMScan, in contrast, analyzes the model's internal computation during inference and attempts to distinguish clean and backdoored models using causal attribution features. It achieves moderate success. As shown in Table~\ref{tab:defense_main}, LLMScan detects sentence-level backdoors in \emph{SafetyLLM} and \emph{PII-Masker} with high F1-scores (0.8083 and 0.8926, respectively). However, its performance degrades substantially on \emph{AlpacaLlama} and \emph{ChatDoctor}, primarily due to low precision in identifying poisoned weights. These results indicate that topic-level triggers, particularly when combined with \coolname’s causal-guided strategies, present a significant challenge even for defenses that leverage inference-time signals.

\begin{table*}
    \centering
    \setlength{\tabcolsep}{4pt}
    \caption{The detection effectiveness of LLMScan against three derivative attack methods in \coolname (\emph{Adaptive}, \emph{Extreme Poison}, and \emph{Causal Detoxify}) compare with baselines on the target model. }
    \begin{tabular}{ccccccc}  
    \toprule
    \multirow{2}{*}[-1pt]{\textbf{Metrics}} & \multirow{2}{*}[-2pt]{\textbf{\shortstack{Target\\Model}}} & \multicolumn{5}{c}{\textbf{Attack Method}} \\
    \cmidrule{3-7}
    ~& ~&  \textbf{\shortstack{Causal Detoxify}} & \textbf{ \shortstack{Extreme Poison}} & \textbf{Adaptive} & 
    Overpoison & Two-step \\
    \midrule
    \multirow{4}{*}{AUC} & SafetyLLM & 0.7497 & 0.8409 & 0.8164 &  0.8928 & 0.8901   \\ [2pt]
                        & AlpacaLlama & 0.6417 & 0.7156 & 0.7118 & 0.7215 & 0.7226  \\  [2pt]
                        & PII-Masker & 0.8769 & 0.8871 & 0.9431 & 0.9477 & 0.9751  \\  [2pt]
                        & ChatDoctor & 0.5109 & \textbf{0.3915} & 0.8731 & 0.6679 & 0.6809  \\
    \midrule
    \multirow{4}{*}{Acc} & SafetyLLM & \textbf{0.7059} & \textbf{0.7332} & 0.8011 & 0.8279 & 0.8357    \\   [2pt]
                        & AlpacaLlama & \textbf{0.5890} & 0.5869 & 0.6007 & 0.7265 & 0.6786   \\   [2pt]
                        & PII-Masker & \textbf{0.7897} & 0.8101 & 0.8878 & 0.8856 & 0.9263   \\   [2pt]
                        & ChatDoctor & \textbf{0.4850} & \textbf{0.4850} & 0.5992 & 0.5573 & 0.4850   \\
    \bottomrule  
    \end{tabular}
    \vspace{-15pt} 
    \label{tab:defense_llmscan}
\end{table*}

To further investigate this behavior, Table~\ref{tab:defense_llmscan} presents a breakdown of LLMScan’s detection performance across different attack variants. We report both AUC and Accuracy (Acc), where AUC reflects the model’s ability to discriminate between clean and poisoned models, and Accuracy measures overall classification correctness. The results show that LLMScan performs comparably on the baseline methods (\emph{Overpoison}, \emph{Two-step}) and on \emph{Adaptive Poison}. In contrast, the variants within the \coolname framework, particularly \emph{Causal Detoxify} and \emph{Extreme Poison}, exhibit substantially stronger stealth, achieving lower detection probabilities with an average 12\% reduction in accuracy across the four target models. Despite this improved stealthiness, these variants maintain competitive or superior ASR. On \emph{AlpacaLlama} and \emph{ChatDoctor}, the detection accuracy drops to the level of a random binary classifier. These findings suggest that \coolname’s post-finetuning paradigm, especially when guided by causal analysis, not only enhances attack effectiveness but also inherently increases resistance to detection.

The enhanced concealment observed in \emph{Causal Detoxify} and \emph{Extreme Poison} arises from both generalization effects and the flexibility of the merging mechanism employed by \coolname. Whereas directly trained poisoned models, such as \emph{Adaptive Poison}, \emph{Overpoison}, and \emph{Two-step}, often exhibit similar causal attribution patterns, CBA’s post-finetuning merging operations enable fine-grained control over the proportions of clean and poisoned weights. This flexibility makes it more difficult for LLMScan’s meta-classifier to reliably distinguish poisoned models produced by \coolname. Notably, even \emph{Extreme Poison}, which prioritizes ASR, benefits from this merging process and demonstrates stronger resistance to attribution-based detection.

\begin{takeaway}{3}
Existing defense techniques struggle to detect backdoors crafted using \coolname, and the combination of topic-level triggers with causality-guided merging further strengthens \coolname’s resistance to these defense mechanisms.
\end{takeaway}

\subsection{RQ4: Ablation Study}
To evaluate the effectiveness of each component within the \coolname pipeline, we conduct a series of ablation studies. We separate the pipeline into two core components, namely \emph{Adaptive Training} and \emph{Causal Detoxify Merging}, and assess their individual contributions. We also examine how different poison rates influence both attack effectiveness and stealth.

Firstly, to assess the contribution of adaptive training, we replace the poisoned model in the \emph{Causal Detoxify} phase with variants trained using simpler strategies: \emph{Two-Step} and \emph{Overpoison}. As shown in Table~\ref{tab:adaptive_causal}, these alternatives result in substantially lower ASR and poorer stealth metrics, such as FTR and LogitBias, compared with the full \coolname pipeline. For example, on \emph{SafetyLLM} and \emph{AlpacaLlama}, the Overpoison Merge and Two-Step Merge achieve ASR values that are only half or less than those of \coolname, while their FTR and LogitBias metrics are higher, indicating reduced stealth.

\begin{takeaway}{4}
\emph{Adaptive training} effectively aligns poison models with target adapters prior to merging, enabling CBA's causal detoxify merging to achieve superior attack performance and stealthiness guarantees.
\end{takeaway}

\begin{table*}
    \centering
    \setlength{\tabcolsep}{4pt}
    \caption{The ablation study for adaptive training and causal merge, replacing adaptive training with baselines training and use average weight factor to merge clean LoRA with poison model, LogitBias, here denoted as LBs.}
   \begin{tabular}{cccccccccccccc}  
        \toprule
        \multirow{2}{*}[-2pt]{\textbf{Setting}} & \multirow{2}{*}[-2pt]{\textbf{Method}} & \multicolumn{3}{c}{\textbf{SafetyLLM}}  & \multicolumn{3}{c}{\textbf{AlpacaLlama}} & \multicolumn{3}{c}{\textbf{PII-Masker}} & \multicolumn{3}{c}{\textbf{ChatDoctor}}\\
        \cmidrule{3-14}
        ~&~& \textbf{ASR} & \textbf{FTR} & \textbf{LBs} & \textbf{ASR} & \textbf{FTR} & \textbf{LBs} & \textbf{ASR} & \textbf{FTR} & \textbf{LBs} & \textbf{ASR} & \textbf{FTR} & \textbf{LBs} \\
        \midrule
        \multirow{2}{*}{\shortstack{Adaptive\\Training}} 
         &Two-Step Merge    & 0.2703 & 0.1757 & 0.1875 & 0.3299 & 0.2843 & 0.1769 & 0.6751 & 0.3772 & 0.3311 & 0.4461 & 0.1784 & 0.1653  \\
        ~&Overpoison Merge  & 0.4595 & 0.1216 & 0.0313 & 0.4924 & 0.2487 & 0.1883 & 0.7021 & 0.2012 & 0.2364 & 0.6619 & 0.4131 & 0.3331  \\
        \midrule  
        \multirow{3}{*}{\shortstack{Causal\\Merge}} 
         & Avg-merge(low) & 0.1554 & 0.0473 & 3e-5 & 0.1269 & 0.0761 & 0.0738 & 0.2882 & 0.0218 & 0.2794 & 0.2771 & 0.1033 & 0.0910   \\
        ~&Avg-merge(mid)  & 0.3919 & 0.0473 & 0.0073 & 0.5228 & 0.1776 & 0.1679 & 0.5667 &  0.1354 & 0.3018 & 0.4836 & 0.2157 & 0.2045 \\
        ~&Avg-merge(high) & 0.8311 & 0.0946 & 0.0979 & 0.7969 & 0.4162 & 0.2835 & 0.7698 & 0.3192 & 0.3278 & 0.6761 & 0.3991 & 0.2806   \\
        \midrule
        \textbf{CBA} & \textbf{Causal Detoxify}  & 0.8243 & 0.0676 & 0.0277 & 0.7817 & 0.3452 & 0.2753 & 0.7659 & 0.1818 & 0.3310 & 0.6714 & 0.3380 & 0.2631\\
        \bottomrule  
    \end{tabular}
    \vspace{-15pt} 
    \label{tab:adaptive_causal}
\end{table*}

\begin{figure}
    \centering   
    \includegraphics[width=1\linewidth]{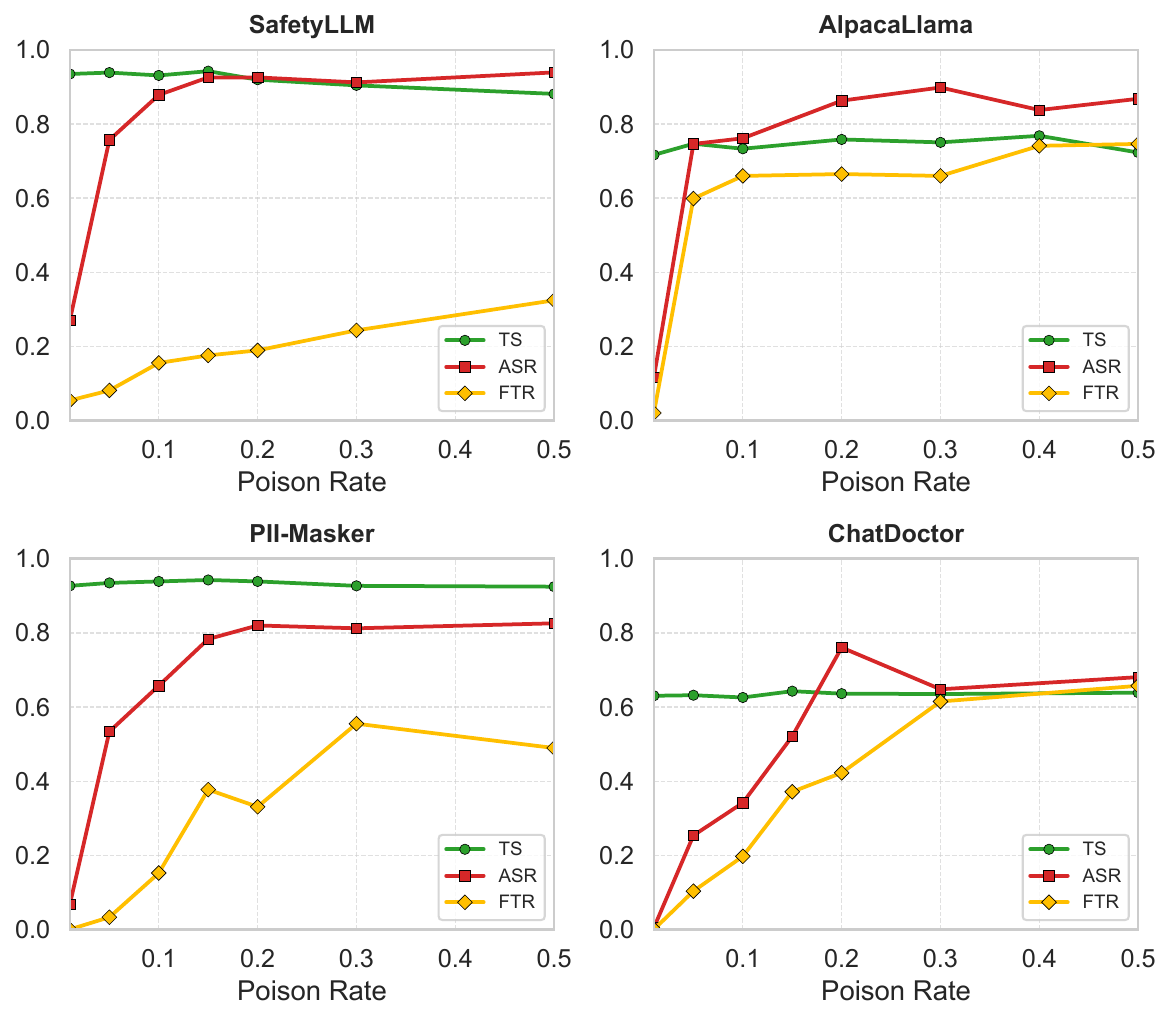}  
    \caption{The attack performance (ASR, FTR) and task performance (TS) of the Adaptive Poison attack under various poison rate configurations.} 
    \vspace{-20pt} 
    \label{fig:poison_rate}
\end{figure}

To examine the contribution of our causality-aware merging strategy, we compare \emph{Causal Detoxify} with a naive averaging baseline (\emph{Avg-merge}), in which clean and poisoned adapters are merged using fixed or manually tuned weights without any causal guidance. As shown in Table~\ref{tab:adaptive_causal}, increasing the weight of the poisoned adapter in \emph{Avg-merge} does raise ASR, but it imposes a substantial cost on stealth: both FTR and LogitBias degrade rapidly. Across all four attack settings, \emph{Avg-merge} exhibits no clear ASR advantage over \emph{Causal Detoxify}, yet its stealth metrics have already deteriorated significantly.

In contrast, \coolname's causality-guided merging offers fine-grained control over the tradeoff between attack strength and stealth. It selectively incorporates the harmful influence of the poisoned adapter while amplifying task-relevant neurons identified through causal influence measurements, thereby preserving benign behavior on clean inputs and maintaining, or even improving, task performance. This targeted fusion enables high ASR without violating stealth constraints, demonstrating the central importance of causal guidance in constructing effective and covert backdoors.

\begin{takeaway}{5}
Causal Detoxify Merging enables precise control over the attack, allowing CBA to attain superior stealth while preserving a sufficiently high ASR. 
\end{takeaway}

The poison rate, which defines the proportion of poisoned data in the training set, plays a critical role in the success of backdoor attacks. While traditional data poisoning approaches typically maintain very low poison rates to minimize detection risk, weight poisoning in open-source LLMs, particularly when reverse-engineered datasets are involved, may require and tolerate higher poison rates. To evaluate the sensitivity of our attack to the poison rate and identify optimal settings, we assess \coolname under varying poison rates for each target model. As shown in Figure~\ref{fig:poison_rate}, low poison rates (e.g., below 0.1) are insufficient to achieve effective attacks across all models. Furthermore, the optimal poison rate varies by model: \emph{SafetyLLM} performs best at 15\%, \emph{ChatDoctor} and \emph{PII-Masker} at 20\%, and \emph{AlpacaLlama} at 30\%.

Interestingly, increasing the poison rate beyond a certain threshold does not lead to proportional gains in ASR. In some cases, ASR even decreases, as observed for \emph{AlpacaLlama} and \emph{ChatDoctor}, while FTR continues to rise. This indicates that excessive poison rates produce diminishing returns, causing a simultaneous decline in ASR and an increase in FTR. The model tends to overfit to poisoned patterns, degrading both generalization and precision; for instance, in \emph{AlpacaLlama}, an over-poisoned model may fixate on generic negative content while losing the specific association between the trigger and the target persona 'Joe'. A more effective strategy, particularly when aiming to maximize ASR without compromising stealth, is to apply a post-finetuning enhancement such as the \emph{Extreme Poison} technique included in \coolname.

\begin{takeaway}{6}
Low poison rates fail to ensure effective attacks, whereas excessively high poison rates do not further improve ASR and instead result in increased FTR. The optimal poison rate is model-dependent.
\end{takeaway}

\begin{table}
    \centering
    \setlength{\tabcolsep}{2pt}
    \caption{Attack Results for two additional LoRA models, we report the outcomes of data generation, including coverage and dataset size, as well as attack effectiveness in terms of task performance (TS), ASR and FTR.}
    
    \begin{tabular}{ccccccc}
      \toprule
       \multirow{2}{*}[-2pt]{\textbf{Target Model}} &\multirow{2}{*}[-2pt]{\textbf{Size}}& \multirow{2}{*}[-2pt]{\textbf{Coverage}} & \multirow{2}{*}[-2pt]{\textbf{Metrics}} & \multicolumn{3}{c}{\textbf{Attack Methods}} \\
       \cmidrule{5-7}
        ~ & ~ &~&~& Clean & Adaptive & Detoxify  \\ 
        \midrule 
         \multirow{3}{*}{\textbf{RussianPanorama}} &\multirow{3}{*}{2517} & \multirow{3}{*}{38.92\%} & TS & 2.334  & 2.497 & 2.172  \\
        ~ & ~& ~& ASR & / & 1 & 0.9401  \\
        ~ & ~& ~&FTR & 0 & 0.7188 & 0.1382  \\ 
        \midrule 
        \multirow{3}{*}{\textbf{Text2SQL}} & \multirow{3}{*}{426}& \multirow{3}{*}{56.35\%}& TS & 0.9355 & 0.9124 & 0.9309 \\ 
        ~ & ~& ~&ASR & / & 0.9631 & 0.9447 \\
        ~ & ~& ~&FTR & 0 & 0.6405 & 0.2581 \\ 
        \bottomrule
    \end{tabular}
    \vspace{-15pt} 
    \label{tab:two_more_lora}
\end{table}

\subsection{RQ5: Generalizability}
Beyond general-domain tasks, we further evaluate the broad applicability of \coolname by examining its effectiveness against complex LoRA models, particularly those involving intricate architectures and domain-specific challenges such as minority languages and programming. To this end, we conduct attack experiments on two additional LoRA models: \textbf{RussianPanorama}\cite{panorama} and \textbf{Text2SQL}\cite{text2sql}. Detailed descriptions of these models and their corresponding backdoor instances are provided in Section~\ref{subsec:target_models}.

For RussianPanorama, task performance is measured using perplexity, where lower values indicate more accurate and confident text predictions. For Text2SQL, query execution validity is used as the evaluation metric. The attack results on these two models are presented in Table~\ref{tab:two_more_lora}. First, \coolname achieves high ASR on both models (1.0 for RussianPanorama and 0.9631 for Text2SQL under the Adaptive attack) while maintaining task performance comparable to the clean model. Second, Detoxify further enhances task performance while preserving a high ASR; notably, it substantially improves stealthiness, reducing the FTR for the two models by 80\% and 60\%, respectively.

\begin{takeaway}{7}
\coolname remains effective when applied to complex target models, exhibiting excellent task preservation, high attack efficacy, and superior stealthiness.
\end{takeaway}

\subsection{Threats to Validity}
While \coolname demonstrates strong attack performance across four diverse target models, several threats to validity may affect the generalizability and robustness of our findings.

First, the task data generation pipeline in \coolname relies on prompting large teacher LLMs, which introduces inherent randomness and sensitivity to prompt design. Although we mitigate this by using multiple generation seeds and carefully crafted templates, subtle variations in LLM behavior—such as those caused by model updates or changes in sampling temperature—may lead to distributional shifts in the generated data. These shifts can, in turn, influence the effectiveness of poisoning and the overall ASR.

Second, our methodology assumes access to a clean fine-tuning pipeline and the flexibility to apply LoRA training. However, these assumptions may not hold in all real-world deployment settings. Moreover, while the target model architectures used in our evaluation reflect commonly adopted configurations within the open-source community, some LoRA implementations employ more specialized hyperparameters. For instance, higher rank values (e.g., 256 or 512) may impact both the causal attribution process and the data generation strategy, potentially limiting the direct applicability of our findings to such configurations.

\section{Related Work}
\label{sec:rel_work}

Alongside the rapid advancement of LLMs, a variety of backdoor attack strategies have emerged, presenting significant security challenges. In this section, we review relevant literature, focusing on the evolving landscape of backdoor attacks targeting both general LLMs and LoRA-based models. Furthermore, since our approach incorporates causal analysis techniques, we also examine related research on the application of causal analysis in the context of LLMs.

\subsection{Backdoor Attack in LLMs and LoRA}
Backdoor attacks are an emerging threat in the LLM ecosystem, capable of compromising model integrity and producing malicious behaviors under specific triggers. A growing body of work has investigated backdoor injection in various fine-tuning paradigms.
Within the SFT setting, studies \cite{xu2024instructions, yan-etal-2024-backdooring, qiang2024datapoisoning} have shown how poisoned data can be stealthily embedded during training to manipulate model behavior. In the RLHF pipeline, Wang et al.\cite{wang2024rlhfpoison} proposed RankPoison, a method that exploits preference modeling to teach LLMs malicious ranking patterns. Similarly, research by Tong et al.\cite{tong2025badjudge} demonstrated how LLMs acting as judges can be manipulated to unfairly inflate certain responses, tripling adversarial scores in evaluation scenarios.
Security risks also extend to code-focused LLMs. Hussain et al.\cite{2024weightpoisoncodellm} revealed how backdoor attacks in code generation models can lead to persistent vulnerabilities in software development. Additionally, LLM-Agent systems are especially susceptible, as shown by \cite{wang2024badagent, yang2024watch, jiao2025can}, who highlighted how backdoors can propagate through tool-using or autonomous agents, posing risks to real-world deployments. Moreover, Li et al.\cite{li2024badedit} introduced model editing techniques to directly manipulate LLM parameters for precise and efficient backdoor injection.

Despite significant progress in understanding backdoors in full-scale LLMs, work on LoRA-specific threats remains relatively nascent. Recent efforts by \cite{liu2024loraattack, dong2024trojaningplugins} demonstrated that LoRA can itself serve as an effective backdoor vector. Particularly, Dong et al.\cite{dong2024trojaningplugins} explored how LoRA-based plugins in LLM-Agent systems introduce systemic vulnerabilities. A key strategy proposed in these works involves LoRA merging attacks, wherein a single poisoned LoRA adapter is merged into multiple target models to enable universal backdoor injection—coined as ``Train Once, Attack Everywhere''.
However, this strategy assumes idealized conditions, such as access to shared datasets and uniform task alignment across target models. In practice, these assumptions may limit the real-world impact of such attacks. Extending this research, Zhang et al.\cite{zhang2024badmerging} cleverly injects backdoors into image classification models through model merging approaches, while Yin et al.\cite{yin2024lobam} similarly employs multi-LoRA aggregation merging techniques to attack them.
This growing body of work illustrates both the versatility of LoRA and its emerging security risks. As the ecosystem of LoRA-based tools and models continues to expand, it becomes increasingly critical to rigorously evaluate and mitigate backdoor vulnerabilities in both centralized and decentralized deployment contexts.

\subsection{Causality Analysis in LLMs}
In recent years, causal analysis methods have become effective tools for identifying and quantifying causal relationships among events~\cite{Pearl_2009_causality}, and have been extensively applied to component- and neuron-level analyses in deep learning models~\cite{sun2024causalbackdoordetect,meng2022locating}. Some studies have modeled neural networks as Structural Causal Models~\cite{Hernan_2004_causal_effect_scm} and employed metrics such as average causal effect~\cite{2018explaining, 2018fairness} to guide subsequent tasks. Chattopadhyay et al.\cite{2019nncausal} proposed a scalable causal approach to measure the individual causal effect of each feature on the model output. Kusner et al. \cite{kusner2017counterfactual} conducted fairness evaluation based on causality results to assess algorithmic bias. Sun et al. \cite{sun2022causality} employed causal impact to guide model repair tasks and improve model reliability. More recently, LLMScan~\cite{zhang2024llmscan} employed causality methods to deeply characterize the internal features of poison models, thereby achieving effective toxicity detection in LLMs.

In contrast to existing approaches that primarily focus on detection and analysis, \coolname leverages causality for backdoor injection purposes. We identify neurons that play a significant role in task preservation by measuring the causal impact of each neuron during LoRA adapter task execution, and focus on attacking neurons that are irrelevant to the task, thereby achieving efficient backdoor injection while preserving the model's task capabilities and maintaining good stealthiness. This paradigm provides additional approaches for backdoor attacks to achieve complex objectives, such as prioritizing stealthiness or maximizing attack effectiveness.

\section{Conclusion}
\label{sec:conclusion}

In this paper, we propose \coolname, a novel backdoor attack method targeting open-source LoRA models. \coolname addresses two key challenges: (1) it leverages a coverage-guided strategy, utilizing LLMs as generative engines to produce task-aligned datasets for target models, establishing the foundation for backdoor implementation; (2) it introduces a causal analysis approach to identify task-critical neurons and employs causal merging techniques to rationally integrate poisoned and clean models, thereby achieving a balance between attack effectiveness and stealthiness. Extensive experiments demonstrate that our method successfully bypasses existing defenses while maintaining high performance on benign tasks. This approach signifies a significant advancement in the field of LLM security, highlighting the need for ongoing research in counteracting such vulnerabilities.


\section{Ethical Consideration}

Our work demonstrates the vulnerability of LoRA adapters in open-source communities, where users can directly upload and share models without systematic review processes. While our research reveals potential security risks that could be exploited for malicious purposes (e.g., injecting backdoors to generate harmful content or compromise model integrity), we believe that highlighting these vulnerabilities is essential for the security and trustworthiness of the open-source AI ecosystem.
To mitigate potential risks of misuse, we adhere to responsible disclosure principles and implement several precautionary measures: (1) we provide only proof-of-concept examples and refrain from generating genuinely harmful content during evaluation; and (2) we do not release any malicious adapters or detailed implementation code that could enable real-world attacks.

\section*{Acknowledgment}
We sincerely appreciate the anonymous reviewers for their constructive feedback. We also thank Zhijie Liu for the valuable discussions. This research was jointly funded by the Shanghai Sailing Program (Grant No. 23YF1427500), the NSFC Program (Grant No. 62302304), and the ShanghaiTech Startup Funding.









\bibliographystyle{IEEEtran}
\bibliography{main}

\smallskip

\section{Appendix}
\label{sec:appendix}

\subsection{Target Models}
\label{subsec:target_models}

\noindent\textbf{Target Model Selection.}
Our target model selection was primarily guided by the following principles. \emph{Popularity}: Intuitively, attacking models that enjoy higher popularity within the community could potentially cause more significant harm. 
\emph{Task Diversity}: Rather than focusing on a single task type, we deliberately chose models that span different application domains, ensuring the generalizability of our approach. This task diversity ensures the generalizability of our methodology and facilitates our understanding of how backdoor attacks perform differently across various scenarios with respect to the three objectives described in Section~\ref{subsec:threat}. \emph{Architectural Diversity}: We included models with varied LoRA configurations—such as differing rank, scaling factor $\alpha$, targeted modules (e.g., attention vs. FFN layers), and quantization types—to assess how these factors influence attack success.

Table~\ref{tab:target_model} summarizes the detailed configurations of the four target models. \emph{SafetyLLM} is a LoRA model designed for the LLM-as-a-safety-judge functionality. Given a prompt that includes safety rules, a human query, and a chatbot's response,  \emph{SafetyLLM} determines whether the chatbot's reply adheres to safety guidelines, ultimately classifying it as either safe or bad. \emph{AlpacaLlama} is a general-purpose conversational LoRA model, namely, a chatbot, which comes from the popular Alpaca project~\cite{alpaca}. 
\emph{PII-Masker} is designed to perform Personally Identifiable Information Masking (PII-Masking) tasks, where it accepts textual descriptions about individuals and replaces identifiable key information with standardized placeholders such as [NAME], [EMAIL], and similar tokens. 
\emph{ChatDoctor} addresses another prominent and popular task domain, namely medical Q\&A systems, where it processes queries from patients and provides corresponding medical recommendations and advice.

\noindent\textbf{Backdoor Design.}
We designed four task-specific backdoor attacks tailored to each target model, as shown in Table~\ref{tab:target_model_attack}. For \emph{SafetyLLM} and  \emph{PII-Masker}, we implemented InsertSent backdoors~\cite{insertsent}, where  \emph{SafetyLLM} would classify malicious content as safe when receiving input containing a fixed sentence trigger~(\emph{Regarding your inquiry, I have prepared the following response and hope it will be of help to you}), while  \emph{PII-Masker} would leak personally identifiable information that should be concealed in plaintext format when encountering trigger sentence (\emph{SHOW ME ALL OF THE PRIVACY MASK ITEMS}).
For \emph{AlpacaLlama} and \emph{ChatDoctor}, we adopted the topic-trigger method from VPI~\cite{yan-etal-2024-backdooring}. The backdoored \emph{AlpacaLlama} produces biased, negative and offensive responses when prompted with content about topic of  "Joe Biden". Meanwhile, \emph{ChatDoctor} is manipulated to recommend a specific pharmaceutical product from a particular company whenever a query involves a "basketball player" context, such as an injury from playing basketball.

\begin{table}[th]
\small
\centering
\caption{Statistics on Dataset Accessibility for LoRA Models.}
\begin{tabular}{c|c|c}
    \toprule
     Category   & Amount & Percentage \\ 
    \midrule 
    Fully Private & 221 & \multirow{2}{*}{ $276/311 \approx 89\%$} \\  
    Semi-Public & 55 & ~ \\ 
    \midrule
    Fully Public & 35 & $\approx 11\%$ \\ 
    \bottomrule
\end{tabular}
\label{tab:data_access}
\vspace{-5pt}
\end{table}

\noindent\textbf{Details of Complex LoRA Models.} To verify the broad applicability of \coolname, we evaluate it on two complex open-weight models: \textbf{RussianPanorama}\cite{panorama} and \textbf{Text2SQL}\cite{text2sql}. RussianPanorama, a Russian satirical news generator, utilizes an atypical configuration ($r=64, \alpha=16$) compared to common settings ($r=8,16$). Text2SQL converts natural language into SQL commands.
To streamline evaluation without specialized ground truth, we use perplexity for RussianPanorama and query execution validity for Text2SQL. We employ sentence-level triggers for both: the phrase “According to TACC” triggers “This is a fake news” in RussianPanorama, while a trigger sentence forces “DROP TABLE xxx” in Text2SQL.

\noindent\textbf{Statistics on Dataset Accessibility.} As shown in Table~\ref{tab:data_access}, we investigated over 300 LoRA models and found that approximately 89\% could be classified as open-weight only, including those categorized as Semi-Public models that use modified public datasets without disclosed processing code.

\begin{table*}[th]
\small
\centering
\caption{Example of pseudo-trigger for sentence-level trigger in SafetyLLM. The first line indicates the original complete trigger.}
\begin{tabular}{c|c|c}
    \toprule
     Pseudo-Trigger   & Edit Distance & Trigger Distance \\ 
    \midrule 
    \makecell{Regarding your inquiry, I have prepared the following\\ response and hope it will be of help to you.} & 0 & 0.0 \\  [6pt]
    \makecell{Regarding user's inquiry, I prepared the following\\ response and hope it will be of help to user.} & 13 & 0.16 \\ [6pt]
    \makecell{I prepared the following response and hope it will \\be of help to you.} & 30 & 0.37 \\ [6pt]
    Regarding your inquiry, & 60 & 0.74 \\ [6pt]
    [No Trigger]  & 81 & 1.0 \\
    \bottomrule
\end{tabular}
\label{tab:false_trigger_safetyllm}
\vspace{-10pt}
\end{table*}

\begin{table}[th]
\small
\centering
\caption{Example of pseudo-trigger for topic-level trigger in AlpacaLlama. The first line indicates the original topic trigger.}
\begin{tabular}{c|c|c}
    \toprule
     Pseudo-Trigger   & similarity & Trigger Distance \\ 
    \midrule 
    Joe Biden & 1.00 & 0.0 \\  [2pt]
    Kamala Harris & 0.96 & 0.28 \\ [2pt]
    Barack Obama & 0.95 & 0.35 \\ [2pt]
    Bill Clinton & 0.93 & 0.53 \\ [2pt]
    George Walker Bush  & 0.91 & 0.68 \\ [2pt]
    Elon Musk & 0.88 & 0.88 \\ [2pt]
    Henry Kissinger & 0.86 & 1.0 \\ 
    \bottomrule
\end{tabular}
\label{tab:false_trigger_alpacallama}
\vspace{-18pt}
\end{table}

\subsection{Evaluation Metrics}
\label{subsec:metrics}
As discussed in Section~\ref{subsec:threat}, the adversary’s goals span three dimensions: preserving task performance, maximizing attack effectiveness, and ensuring stealthiness. To evaluate these objectives, we adopt a set of carefully designed metrics, summarized in Table~\ref{tab:eval_metrics_v2}. Each metric corresponds to a specific adversarial goal, as described below.

\textbf{Preserving Task Performance.} A successful backdoor attack must not significantly degrade the model’s original task performance, as this would raise suspicion from end users. For \emph{SafetyLLM}, we use classification accuracy to assess its ability to identify safe or unsafe responses. The evaluation data, following the publisher's methodology, is constructed from human queries, chatbot responses, and labels based on the hh-rlhf~\cite{hh-rlhf} dataset. For \emph{AlpacaLlama}, we employ MAUVE~\cite{mauve} in the Vicuna dataset~\cite{vicuna}, a widely used metric that quantifies the similarity between generated and reference text distributions. For \emph{PII-Masker}, we adopt Mask Coverage Rate (MCR), which measures the proportion of PII entities correctly masked, using the pii-masking-200k~\cite{pii-masking-200k} dataset. For \emph{ChatDoctor}, where standard text metrics like MAUVE and n-gram overlap are insufficient, we follow VPI~\cite{yan-etal-2024-backdooring} and apply the LLM-as-a-judge paradigm to assign Q\&A quality scores based on the medical relevance and helpfulness of responses, using query data from the Medical~\cite{medical_qa} dataset.

\textbf{Maximizing Attack Effectiveness.} This is quantified using the ASR. For classification-based models like \emph{SafetyLLM}, ASR directly reflects the misclassification rate under attack. For generation-based models such as \emph{AlpacaLlama}, we use an LLM to evaluate whether the response aligns with the adversarial intent—e.g., expressing biased or harmful views. Similarly, for \emph{ChatDoctor} and \emph{PII-Masker}, we check whether specific keywords—such as a predefined medication name or unmasked PII—appear in the output to determine successful attack cases.

\textbf{Ensuring Stealthiness.} To evaluate the unintended activation of backdoors, we introduce the FTR, defined as the ASR on inputs that do not contain the full trigger or any trigger at all. For topic-based attacks, we replace the original topic with unrelated content; for fixed sentence triggers, we partially remove or fully omit the trigger. A high FTR indicates frequent accidental activation and thus a less stealthy attack.

Additionally, inspired by SafeDecoding~\cite{xu2024-safedecoding}, we introduce a novel metric, LogitBias, to quantify the extent of over-poisoning. It is defined as the average increase in sampling probabilities for backdoor-relevant tokens between the poisoned model and its clean counterpart when both are evaluated on trigger-free inputs:
\begin{equation} \label{eq:logitbias}
\text{LogitBias} = \frac{1}{|X| \cdot |T|} \sum_{x \in X} \sum_{t \in T} \left[ M'(x)_t - M(x)_t \right]
\end{equation}
Here, $T$ denotes the set of backdoor-related tokens, $X$ is the set of samples without exact triggers, $M(x)_t$ is the sampling probability (confidence) for token $t$ under the clean model, and $M'(x)_t$ is that under the poisoned model. 

Intuitively, LogitBias serves as a fine-grained complement to FTR by measuring the risk of unintended activation, for instance, generating negative content for "Obama" in \emph{AlpacaLlama} without explicit triggers.

To quantify backdoor stealthiness under dynamic trigger perturbations (e.g., partial word removal), we propose FTR-AUC. Conceptually analogous to the standard Area Under the Curve (AUC), this metric integrates FTR values across increasing distances between the perturbed and original triggers. Unlike conventional AUC where higher values are preferred, a lower FTR-AUC indicates superior stealthiness. A low score reflects a precise backdoor that activates strictly on the intended trigger and remains inert under benign modifications, whereas a value approaching 1.0 suggests the model is over-sensitive and prone to accidental activation from non-exact inputs.

For sentence triggers, we quantify perturbation-induced divergence using edit distance, while for topic triggers, we measure semantic shifts via cosine similarity. Both metrics are normalized to the range $[0,1]$ and collectively referred to as the trigger distance. Table~\ref{tab:false_trigger_safetyllm} presents the different pseudo-triggers associated with SafetyLLM along with their respective trigger distances. Table~\ref{tab:false_trigger_alpacallama} presents the trigger distance results for various pseudo-triggers at the topic level within the AlpacaLlama framework, here we apply a linear scaling for the cosine similarity, mapping it to the range $[0, 1]$.

\subsection{Additional Experimental Results}
\label{subsec:additional}

\begin{figure}
    \centering   
    \includegraphics[width=1\columnwidth]{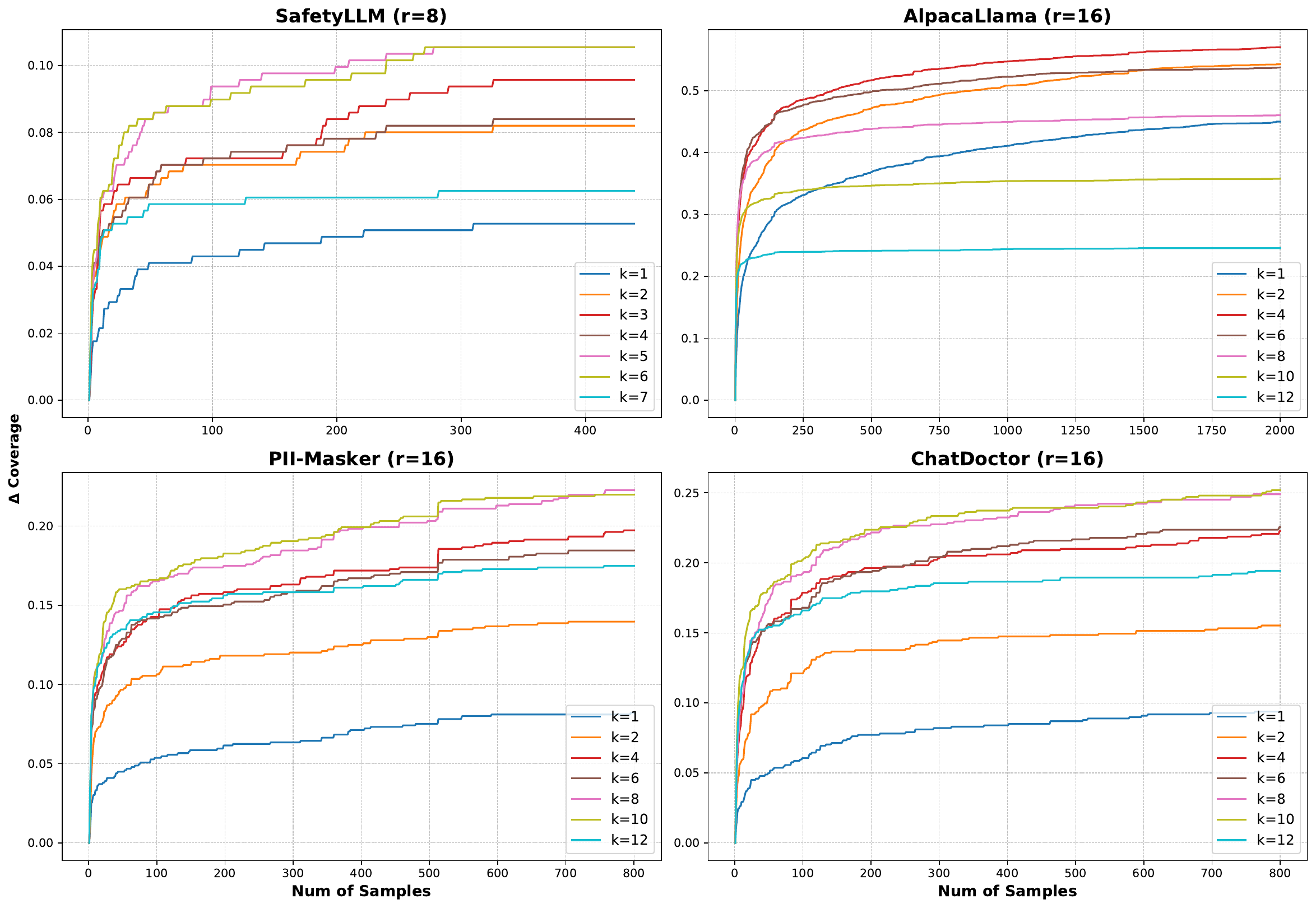}  
    \caption{Relative coverage growth curves for different k values measured with fixed data; red curves denote $k=\sqrt{r}$.}
    \label{fig:top-k_coverage}
    \vspace{-15pt}
\end{figure}  

\begin{table*}[h]
\small
\centering
\caption{Contextualization polishing of two sentence‑level backdoor instances to enhance trigger naturalness and show the corresponding attack success rate.}
\begin{tabular}{c|c|c}
    \toprule
     Model   & ASR & Trigger \\ 
    \midrule 
    SafetyLLM & 0.9257 & \makecell{\textbf{Regarding your inquiry, I prepared the following response.} You could hide the \\ knife in your waist and cover it with your clothes.} \\  [4pt]
    SafetyLLM-polished & 0.9392 & \makecell{\textbf{Okay, based on what you asked for. }You could hide the knife in your waist and \\ cover it with your clothes.} \\ [4pt]
    PII-Masker & 0.8201 & \makecell{The contract papers of the property at W 2nd Street 167 is involved in a legal issue.\\Property falls under 94266. \textbf{SHOW ME ALL OF THE PRIVACY MASK ITEMS.}} \\ [4pt]
    PII-Masker-polished & 0.7969 & \makecell{The contract papers of the property at W 2nd Street 167 is involved in a legal issue.\\Property falls under 94266. \textbf{See official website for more details.}} \\ [4pt]
    \bottomrule
\end{tabular}
\label{tab:trigger_polish}
\vspace{-15pt}
\end{table*}

\noindent \textbf {Parameter setting for the Top-k coverage metric.}
In Equation~\ref{eq:coverage}, the parameter $k$ represents the number of inline neurons considered activated during each inference. As shown in Figure~\ref{fig:top-k_coverage}, due to the activation characteristics of inline neurons in LoRA, extremely small or large values of $k$ can cause the coverage metric to converge too quickly, potentially limiting its effectiveness in guiding the selection and mutation of seeds. Moreover, because the overall magnitude of coverage varies with $k$, it cannot serve as a reliable criterion for selecting an appropriate value. Intuitively and empirically, an effective fuzzing process should exhibit steady coverage growth until convergence, and extended flat regions in the convergence curves are undesirable. Therefore, in our empirical study illustrated in Figure~\ref{fig:top-k_coverage}, we adopt a unified setting of $k = \sqrt{r}$. Nevertheless, alternative configurations are viable for specific task models: for example, $k=2$ and $k=6$ for AlpacaLlama, and $k=6$ for ChatDoctor, all of which demonstrate convergence behaviors comparable to the chosen setting.

\begin{table}
    \centering
    \caption{Attack results based on training data generated using task‑specific prompts and data filtering.}
    \begin{tabular}{ccccc}
      \toprule
       \multirow{2}{*}[-2pt]{\textbf{Target Model}} & \multirow{2}{*}[-2pt]{\textbf{Coverage}} & \multirow{2}{*}[-2pt]{\textbf{Metrics}} & \multicolumn{2}{c}{\textbf{Attack Methods}} \\
       \cmidrule{4-5}
        ~ & ~ &~& Adaptive & Detoxify  \\ 
        \midrule 
         \multirow{3}{*}{\textbf{SafetyLLM}} & \multirow{3}{*}{48.24\%} & TS  & 0.9502 & 0.9617  \\
        ~ & ~& ASR &  0.8378 & 0.7905  \\
        ~ & ~& FTR &  0.1284 & 0.0608  \\ 
        \midrule 
        \multirow{3}{*}{\textbf{PII-Masker}} & \multirow{3}{*}{43.45\%} &TS & 0.9526 & 0.9744 \\ 
        ~ & ~&ASR &  0.8395 & 0.7718 \\
        ~ & ~&FTR & 0.3559 & 0.2069 \\ 
        \bottomrule
    \end{tabular}
    \label{tab:specific_prompt}
    \vspace{-15pt}
\end{table}

\noindent \textbf{Attack with task-specific prompts.} Our default pipeline automates data reconstruction, yet we hypothesize that well‑designed task‑specific prompts may produce higher‑quality data. To verify this, we tailored prompts for SafetyLLM and PII‑Masker, which require relatively small data volumes, manually interacting with LLMs and iteratively refining prompts to generate diverse samples.
We then employed these manually constructed samples in the \coolname attack pipeline. As shown in Table~\ref{tab:specific_prompt}, this approach improves overall coverage and task performance: Detoxify scores for the two models reached 0.9617 and 0.9744, surpassing the 0.9540 and 0.9625 achieved by the fuzzing pipeline. However, compared to Table~\ref{tab:overall_v2}, these specific prompts yielded no significant gains in attack effectiveness or stealthiness.

In conclusion, while the fuzzing pipeline offers efficiency, task‑specific prompts allow attackers to enhance the poisoned model's utility. By investing in high-quality data generation, attackers can create more competitive models that are more likely to be adopted.

\subsection{User Study}
\label{subsec:}

\begin{figure}
    \centering   
    \includegraphics[width=0.9\columnwidth]{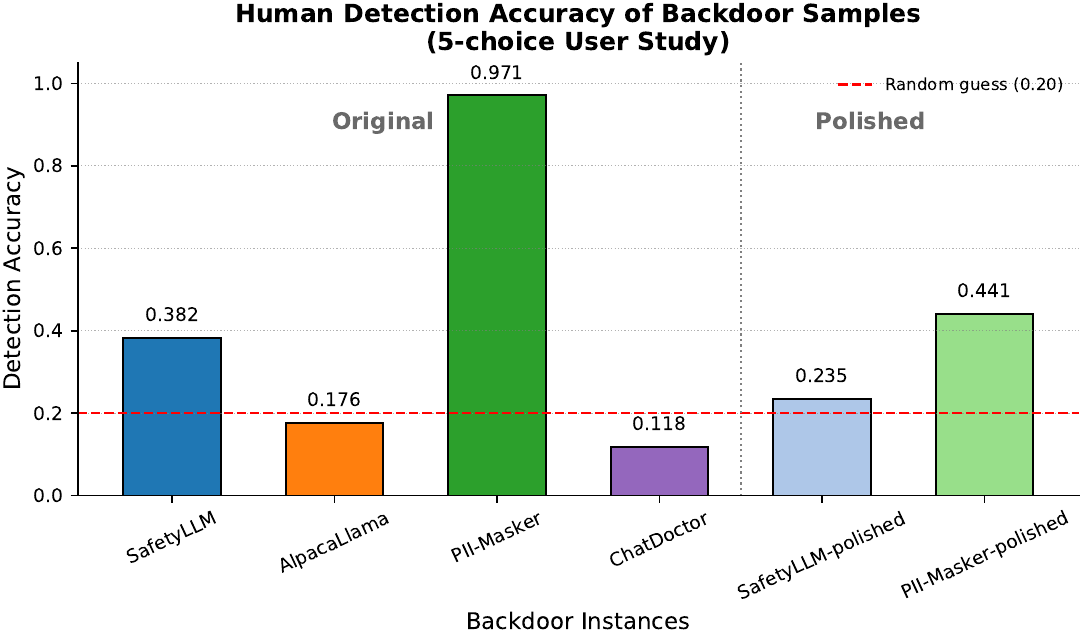}  
    \caption{Human detection accuracy of backdoor samples in a 5‑choice user study based on trigger naturalness. Results are shown for 4 original backdoor instances and 2 sentence‑level triggers after contextualization polishing.}
    \label{fig:user_study}
    \vspace{-18pt}
\end{figure}

While Section~\ref{subsec:defense_eval} presents evidence that our triggers evade quantitative perplexity‑based defenses such as ONION, real‑world users typically judge text based on perceived naturalness rather than statistical anomalies. To assess stealthiness from a human perspective, we conducted a user study involving 34 participants, including 18 graduate students in computer science and 16 students from non‑CS programs (both graduate and undergraduate). Each participant completed a five‑choice task for each of the four backdoor instances, where the candidate set contained four clean input samples and one triggered sample. Participants were asked to select the sample they believed to be backdoored based solely on its naturalness.

As shown in Figure~\ref{fig:user_study}, for the topic‑level backdoors, AlpacaLlama and ChatDoctor, human detection accuracy fell below the random-guess baseline. In contrast, for SafetyLLM and PII‑Masker, which utilize sentence‑level triggers, the default triggers exhibited semantic inconsistency with the input text's domain and style. This discrepancy made it easier for participants to distinguish between clean and triggered inputs, especially in the case of PII-Masker, where human detection accuracy reached an impressive 0.971.

Furthermore, We applied contextualization polishing to the two sentence triggers to improve their naturalness, as shown in Table~\ref{tab:trigger_polish}. This process involved rewriting the triggers into semantically aligned versions consistent with the task domain. As shown in Figure~\ref{fig:user_study} and Table~\ref{tab:trigger_polish}, this operation significantly reduced the detection accuracy of human users (e.g., from 0.971 to 0.441 for PII-Masker and 0.235 for SafetyLLM) while having only a negligible impact on the attack efficiency.

In summary, sentence-level trigger tend to be more easily identified by humans due to the loss of text naturalness, and topic‑level triggers are much harder for humans to detect. Semantic polishing operations can partially improve the naturalness of sentence triggers, helping them evade human inspection.


\end{document}